# Audio-Visual Speech Enhancement Using Multimodal Deep Convolutional Neural Networks

Jen-Cheng Hou, Syu-Siang Wang, Ying-Hui Lai, Yu Tsao, *Member, IEEE*, Hsiu-Wen Chang, and Hsin-Min Wang, *Senior Member, IEEE*

***Abstract*—Speech enhancement (SE) aims to reduce noise in speech signals. Most SE techniques focus only on addressing audio information. In this work, inspired by multimodal learning, which utilizes data from different modalities, and the recent success of convolutional neural networks (CNNs) in SE, we propose an audio-visual deep CNNs (AVDCNN) SE model, which incorporates audio and visual streams into a unified network model. We also propose a multi-task learning framework for reconstructing audio and visual signals at the output layer. Precisely speaking, the proposed AVDCNN model is structured as an audio-visual encoder-decoder network, in which audio and visual data are first processed using individual CNNs, and then fused into a joint network to generate enhanced speech (the primary task) and reconstructed images (the secondary task) at the output layer. The model is trained in an end-to-end manner, and parameters are jointly learned through back-propagation. We evaluate enhanced speech using five instrumental criteria. Results show that the AVDCNN model yields a notably superior performance compared with an audio-only CNN-based SE model and two conventional SE approaches, confirming the effectiveness of integrating visual information into the SE process. In addition, the AVDCNN model also outperforms an existing audio-visual SE model, confirming its capability of effectively combining audio and visual information in SE.**

***Index Terms*—Audio-visual systems, deep convolutional neural networks, multimodal learning, speech enhancement.**

## I. Introduction

The primary goal of speech enhancement (SE) is to enhance the intelligibility and quality of noisy speech signals by reducing the noise components of noise-corrupted speech. To attain a satisfactory performance, SE has been used as a fundamental unit in various speech-related applications, such as automatic speech recognition [1, 2], speaker recognition [3, 4], speech coding [5, 6], hearing aids [7–9], and cochlear implants [10–12]. In the past few decades, numerous SE methods have been proposed and proven to provide an improved sound quality. One notable approach, spectral restoration, estimates a gain function (based on the statistics of noise and speech components), which is then used to suppress noise components in the frequency domain to obtain a clean speech spectrum from a noisy speech one [13–17]. Another class of approaches adopts a nonlinear model to map noisy speech signals to clean ones [18–21]. In recent years, SE methods based on deep learning have been proposed and investigated extensively, such as denoising autoencoders [22, 23]. SE methods using deep neural networks (DNNs) generally exhibit better performances than conventional SE models [24–26]. Approaches that utilize recurrent neural networks (RNNs) and long short-term memory (LSTM) models have also been confirmed to exhibit promising SE and related speech signal processing performances [27–29]. In addition, inspired by the success of image recognition using convolutional neural networks (CNNs), a CNN-based model has been shown to obtain good results in SE owing to its strength in handling image-like 2-D time-frequency representations of noisy speech [31, 32].

In addition to speech signals, visual information is important in human-human or human-machine interactions. A study of the McGurk effect [33] indicated that the shape of the mouth or lips could play an important role in speech processing. Accordingly, audio-visual multimodality has been adopted in numerous fields of speech-processing [34–39]. The results have shown that visual modality enhances the performance of speech processing compared with its counterpart that uses audio modality alone. In addition, topics regarding the fusion of audio and visual features have been addressed in [40, 41], where additional reliability measures were adopted to achieve a better dynamic weighting of audio and visual streams. On the other hand, in [42, 43] intuitive fusion schemes were adopted in multimodal learning based on the architectures of neural networks. There have also been several related studies in the field of audio-visual SE [44–50]. Most of these are based on an enhancement filter, with the help of handcrafted visual features from lip

Jen-Cheng Hou is with the Research Center for Information Technology Innovation at Academia Sinica, Taipei, Taiwan. (email:coolkiu@citi.sinica.edu.tw).

Syu-Siang Wang is with the Graduate Institute of Communication Engineering, National Taiwan University, Taipei, Taiwan. (email:d02942007@ntu.edu.tw).

Ying-Hui Lai is with Department of Biomedical Engineering, National Yang-Ming University, Taipei, Taiwan. (email: yh.lai@ym.edu.tw).

Yu Tsao is with the Research Center for Information Technology Innovation at Academia Sinica, Taipei, Taiwan. (email:yu.tsao@citi.sinica.edu.tw).

Hsiu-Wen Chang is with Department of Audiology and Speech language pathology, Mackay Medical College, New Taipei City, Taiwan. (email:hsiuwen@mmc.edu.tw).

Hsin-Min Wang is with the Institute of Information Science at Academia Sinica, Taipei, Taiwan. (email:whm@iis.sinica.edu.tw).

shape information. Recently, some audio-visual SE models based on deep learning have also been proposed [51, 52]. In [51], Mel filter banks and a Gauss-Newton deformable part model [53] were used to extract audio and mouth shape features. Experimental results showed that DNNs with audio-visual inputs outperformed DNNs with only audio inputs in several standardized instrumental evaluations. In [52], the authors proposed dealing with audio and visual data using DNNs and CNNs, respectively. The noisy audio features and the corresponding video features were used as input, and the audio features were used as the target during training.

In the present work, we adopted CNNs to process both audio and visual streams. The outputs of the two networks were fused into a joint network. Noisy speech and visual data were placed at inputs, and clean speech and visual data were placed at outputs. The entire model was trained in an end-to-end manner, and structured as an audio-visual encoder-decoder network. Notably, the visual information at the output layer served as part of the constraints during the training of the model, and thus the system adopted a multi-task learning scheme that considered heterogeneous information. Such a unique audio-visual encoder-decoder network design has not been used in any related work [51, 52]. In short, the proposed audio-visual SE model takes advantage of CNNs, which have shown to be effective in speech enhancement [31, 32] and image and face recognition [54, 55], for both audio and visual streams, and the properties of deep learning, i.e., reducing human-engineering efforts by end-to-end learning and intuitive fusion schemes in multimodal learning tasks. To our best knowledge, this is the first model to exploit all of the aforementioned properties at once in a deep learning-based audio-visual SE model.

Our experimental results show that the proposed audio-visual SE model outperforms four baseline models, including three audio-only SE models and the audio-visual SE model in [51], in terms of several standard evaluation metrics, including the perceptual evaluation of speech quality (PESQ) [56], short-time objective intelligibility (STOI) [57], speech distortion index (SDI) [58], hearing-aid speech quality index (HASQI) [59], and hearing-aid speech perception index (HASPI) [60]. This confirms the effectiveness of incorporating visual information into the CNN-based multimodal SE framework, and its superior efficacy in combining heterogeneous information as an audio-visual SE model. In addition, an alternative fusion scheme (i.e., early fusion) based on our audio-visual model is also evaluated, and the results show that the proposed architecture is superior to the early fusion one.

The remainder of this paper is organized as follows. Section II describes the preprocessing of audio and visual streams. Section III introduces the proposed CNN-based audio-visual model for SE, and describes the four baseline models for comparison. Section IV describes the experimental setup and results, and a discussion follows in Section V. Section VI presents the concluding remarks of this study.

## II. Dataset and Preprocessing

In this section, we provide the details of the datasets and preprocessing steps for audio and visual streams.

### A. Data Collection

The prepared dataset contains video recordings of 320 utterances of Mandarin sentences spoken by a native speaker. The script for recording is based on the Taiwan Mandarin hearing in noise test (Taiwan MHINT) [61], which contains 16 lists, each including 20 sentences. The sentences are specially designed to have similar phonemic characteristics among lists. Each sentence is unique and contains 10 Chinese characters. The length of each utterance is approximately 3–4 seconds. The utterances were recorded in a quiet room with sufficient light, and the speaker was filmed from the front view. Videos were recorded at 30 frames per second (fps), at a resolution of 1920 pixels × 1080 pixels. Stereo audio channels were recorded at 48 kHz. 280 utterances were randomly selected as a training set, with the remaining 40 utterances used as the testing set.

### B. Audio Feature Extraction

We resampled the audio signal to 16 kHz, and used only a mono channel for further processing. Speech signals were converted into the frequency domain and processed into a sequence of frames using the short-time Fourier transform. Each frame contained a window of 32 milliseconds, and the window overlap ratio was 37.5%. Therefore, there were 50 frames per second. For each speech frame, we extracted the logarithmic power spectrum, and normalized the value by removing the mean and dividing by the standard deviation. The normalization process was conducted at the utterance level, i.e., the mean and standard deviation vectors were calculated from all frames of an utterance. We concatenated ±2 frames to the central frame as context windows. Accordingly, audio features had dimensions of 257 × 5 at each time step. We use $X$ and $Y$ to denote noisy and clean speech features, respectively.

### C. Visual Feature Extraction

For the visual stream, we converted each video that contained an utterance into an image sequence at a fixed frame rate of 50 fps, in order to keep synchronization of the speech frames and the image frames. Next, we detected the mouth using the Viola–Jones method [62], resized the cropped mouth region to 16 pixels × 24 pixels, and retained its RGB channels. In each channel, we rescaled the image pixel intensities in a range of 0 to 1. We subtracted the mean and divided it by the standard deviation for normalization. This normalization was conducted for each colored mouth image. In addition, we concatenated ±2 frames to the central frame, resulting in visual features with dimensions of 16 × 24 × 3 × 5 at each time step. We use $Z$ to represent input visual features.

For each utterance, the number of frames of audio spectrogram and the number of mouth images were made the same using truncation if necessary.

## III. Audio-Visual Deep Convolutional Neural Networks (AVDCNN)

The architecture of the proposed AVDCNN model is illustrated in Fig. 1. It is composed of two individual networks that handle audio and visual steams, respectively, namely the audio network and visual network. The outputs of the two networks are fused into another network, called the fusion network. The CNN, maximum pooling layer, and fully-connected layer in the diagram are abbreviated as $Conv_a1$, $Conv_a2$, $Conv_v1$, …, $Pool_a1$, FC1, FC2, $FC_a3$, and $FC_v3$, where the subscripts 'a' and 'v' denote the audio and visual stream, respectively. In the following section, we describe the training procedure of the AVDCNN model.

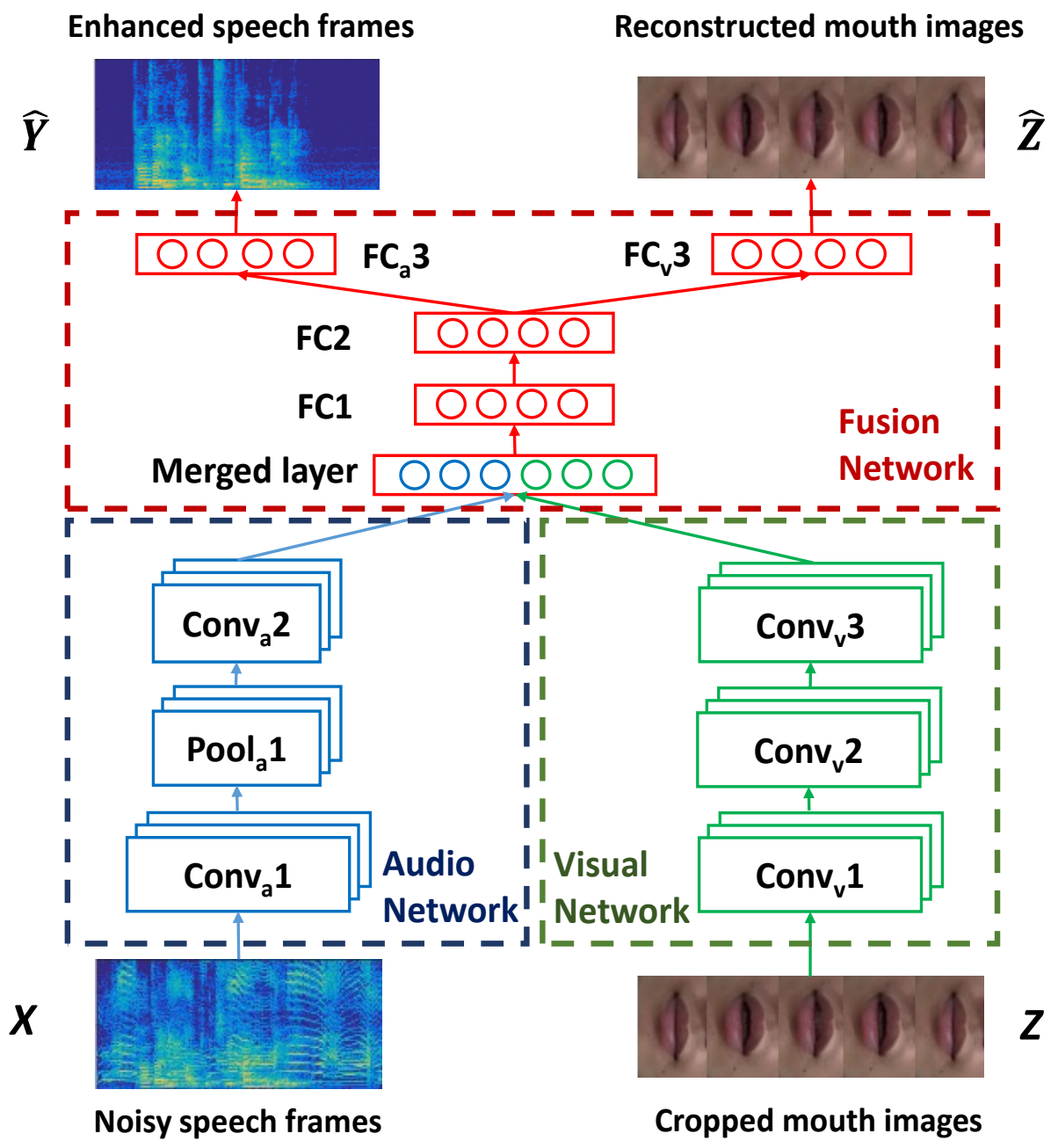

Fig. 1. The architecture of the proposed AVDCNN model.

### A. *Training the AVDCNN Model*

To train the AVDCNN model, we first prepare noisy-clean speech pairs and mouth images. As described in parts B and C in Section II, we have the logarithmic amplitudes of noisy ($X$) and clean ($Y$) spectra and the corresponding visual features ($Z$). For each time step, we obtain the output of the audio network as

$$A_i = Conv_a2\left(Pool_a1\left(Conv_a1(X_i)\right)\right), i = 1 \dots K \quad (1)$$

where $K$ is the number of training samples. The output of the visual network is

$$V_i = Conv_v3\left(Conv_v2\left(Conv_v1(Z_i)\right)\right), i = 1 \dots K \quad (2)$$

Next, we flatten $A_i$ and $V_i$, and concatenate the two features as the input of the fusion network, $F_i = [A_i{}' \ V_i{}']'$. A feed-forward cascaded fully-connected network is computed as:

$$\hat{Y}_i = FC_a3\left(FC2\left(FC1(F_i)\right)\right), i = 1 \dots K \quad (3)$$

$$\hat{Z}_i = FC_v3\left(FC2\left(FC1(F_i)\right)\right), i = 1 \dots K \quad (4)$$

The parameters of the AVDCNN model, denoted as $\theta$, are randomly initialized between -1 and 1, and are trained by optimizing the following objective function using back-propagation:

$$\min_\theta \left(\frac{1}{K}\sum_{i=1}^{K}\left\|\hat{Y}_i - Y_i\right\|_2^2 + \mu\left\|\hat{Z}_i - Z_i\right\|_2^2\right), \quad (5)$$

where $\mu$ is a mixing weight.

A stride size of 1 × 1 is adopted in the CNNs of the AVDCNN model, and a dropout of 0.1 is adopted after FC1 and FC2 to prevent overfitting. Batch normalization is applied for each layer in the model. Other configuration details are presented in Table I.

### B. *Using the AVDCNN Model for Speech Enhancement*

In the testing phase, the logarithmic amplitudes of noisy speech signals and the corresponding visual features are fed into the trained AVDCNN model to obtain the logarithmic amplitudes of enhanced speech signals and corresponding visual features as outputs. Similar to spectral restoration approaches, the phases of the noisy speech are borrowed as the phases for the enhanced speech. Then, the AVDCNN-enhanced amplitudes and phase information are used to synthesize the enhanced speech. We consider the visual features at the output of the trained AVDCNN model only as auxiliary information. This special design enables the AVDCNN model to process audio and visual information concurrently. Thus, the training process is performed in a multi-task learning manner, which has been proven to achieve a better performance than single-task learning in several tasks [63, 64].

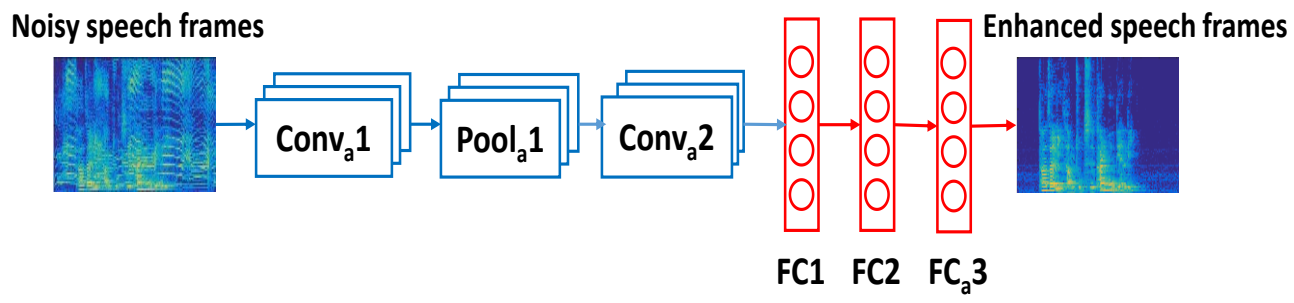

Fig. 2. The architecture of the ADCNN model, which is the same as the AVDCNN model in Fig. 1 with the visual parts disconnected.

TABLE I
CONFIGURATIONS OF THE AVDCNN MODEL

| Layer Name | Kernel | Activation Function | Number of Filters or Neurons |
|---|---|---|---|
| $Conv_a1$ | 12 × 2 | Linear | 10 |
| $Pool_a1$ | 2 × 1 | | |
| $Conv_a2$ | 5 × 1 | Linear | 4 |
| $Conv_v1$ | 15 × 2 | Linear | 12 |
| $Conv_v2$ | 7 × 2 | Linear | 10 |
| $Conv_v3$ | 3 × 2 | Linear | 6 |
| Merged Layer | | | 2804 |
| FC1 | | Sigmoid | 1000 |
| FC2 | | Sigmoid | 800 |
| $FC_a3$ | | Linear | 600 |
| $FC_v3$ | | Linear | 1500 |

### C. *Baseline Models*

In this work, we compare the proposed AVDCNN model with three audio-only baseline models. The first is the audio-only deep CNNs (ADCNN) model. As shown in Fig. 2, the ADCNN model disconnects all visual-related parts in the AVDCNN model (cf. Fig. 1), and keeps the remaining configurations. The second and third are two conventional SE approaches, namely the Karhunen-Loéve transform (KLT) [65] and log minimum mean squared error (logMMSE) [66, 67].

In addition, the audio-visual SE model in [51], denoted by AVDNN, is adopted as an audio-visual baseline model. The AVDNN model employs handcrafted audio and visual features, consisting of Mel-filter banks and the mutual distance changes between points in the lip contour, respectively. Another main difference between AVDCNN and AVDNN is that the AVDNN model is based on DNNs and does not adopt the multi-tasking learning scheme, while AVDCNN applies multi-task learning by considering audio and visual information at the output layer.

## IV. Experiments and Results

### A. *Experimental Setup*

In this section, we describe the experimental setup for the speech enhancement task in this study. To prepare the clean-noisy speech pairs, we follow the concept in the previous study [68], where the effects of both interference noise and ambient noise were considered. For the training set, we used 91 different noise types as interference noises. These 91 noises were a subset of the 104 noise types used in [69, 70]. Thirteen noise types that were similar to the test noise types were removed. Car engine noises under five driving conditions were used to form the ambient noise set, including idle engine, 35 mph with the windows up, 35 mph with the windows down, 55 mph with the windows up, and 55 mph with the windows down. The car engine noises were taken from the AVICAR dataset [71]. We concatenated these car noises to form our final ambient noise source for training. To form the training set, we first randomly chose 280 out of 320 clean utterances. The clean utterances were artificially mixed with 91 noise types at 10 dB, 6 dB, 2 dB, -2 dB, and -6 dB signal-to-interference noise ratios (SIRs), and the ambient noise at 10 dB, 6 dB, 2 dB, -2 dB, and -6 dB signal-to-ambient noise ratios (SARs), resulting in a total of (280×91×5×5) utterances.

Next, to form the testing set we adopted 10 types of interference noises, including a baby crying sound, pure music, music with lyrics, a siren, one background talker (1T), two background talkers (2T), and three background talkers (3T), where for the 1T, 2T, and 3T background talker noises there were two modes: on-air recording and room recording. These noises were unseen in the training set, i.e., a noise-mismatched condition was adopted. Furthermore, they were chosen in particular because we intended to simulate a car driving condition as our test scenario, such as listening to the radio while driving with noises from talkers in the rear seats and the car engine, given that audio-visual signal processing techniques had been effective in improving in-car voice command systems [72–74]. In addition, the ambient noise for testing was a 60 mph car engine noise taken from the dataset used in [75], which was also different from those used in the training set. Consequently, for testing there were 40 clean utterances, mixed with the 10 noise types at 5 dB, 0 dB, and -5 dB SIRs, and one ambient noise at 5 dB, 0 dB, and -5 dB SARs, resulting in a total of (40×10×3×3) utterances.

We used stochastic gradient descent and RMSprop [76] as the learning optimizer to train the neural network model, with an initial learning rate of 0.0001. We chose the weights of the model where the following 20 epochs exhibited improvements of less than 0.1% in the training loss. The implementation was based on the Keras [77] library.

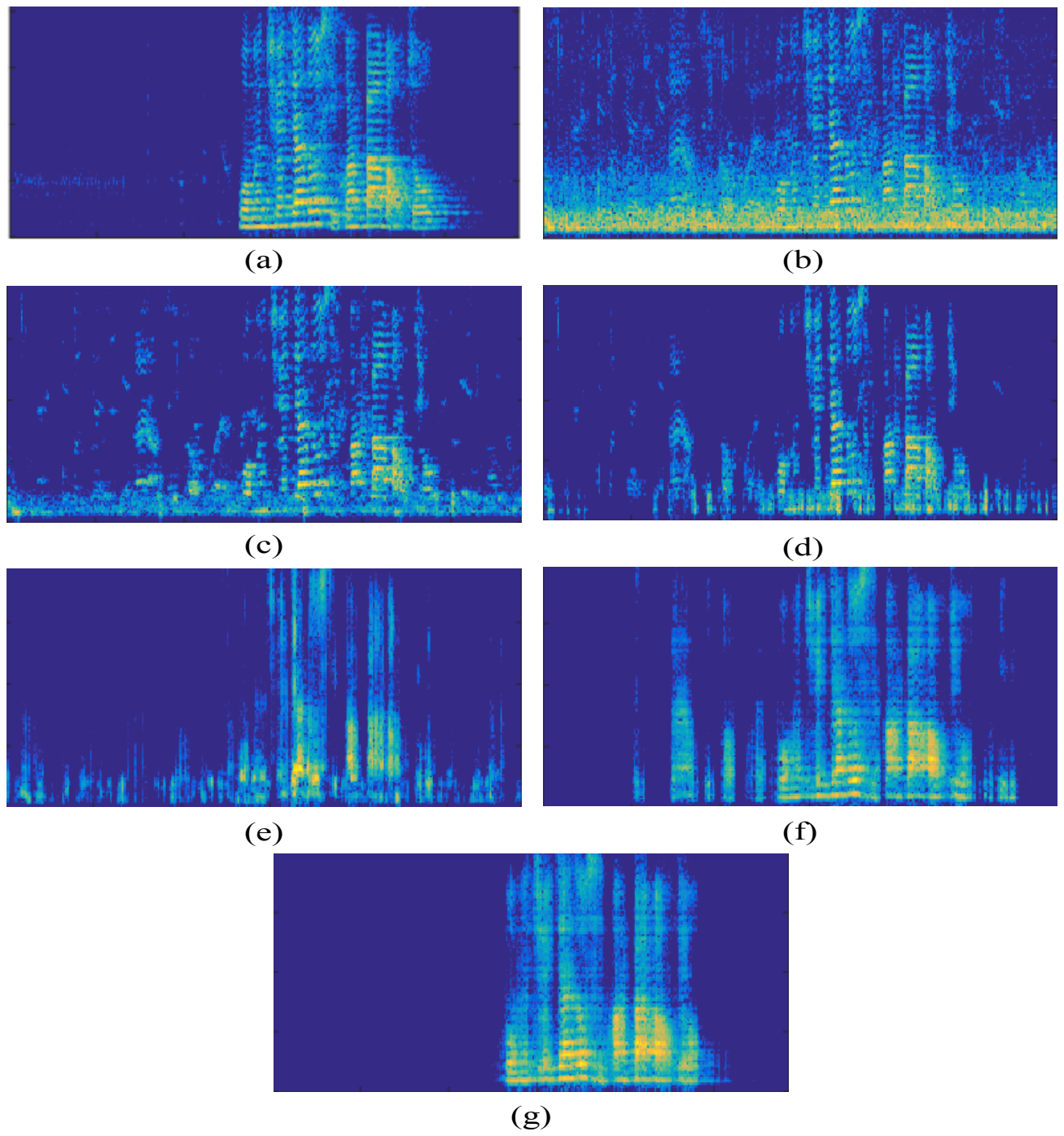

Fig. 3 Comparison of spectrograms: (a) the clean speech, (b) the noisy speech of the 3T (room) noise at 5 dB SIR with the ambient noise at -5 dB SAR, and the speech enhanced by (c) logMMSE, (d) KLT, (e) AVDNN, (f) ADCNN, and (g) AVDCNN.

### B. *Comparison of Spectrograms*

Fig. 3, (a)-(g) present the spectrograms of clean speech, noisy speech mixed with 3T (room) noise at 5 dB SIR with -5 dB SAR, and speech enhanced by the logMMSE, KLT, AVDNN, ADCNN, and AVDCNN methods, respectively. It is obvious that all three audio-only SE approaches could not effectively remove the noise components. This phenomenon is especially clear for the silence portions at the beginning and the end of the utterance, where noise components can still be observed. In contrast, with the help of auxiliary visual information, the AVDCNN model effectively suppressed the noise components in the parts where the mouth was closed.

However, even with additional visual information the AVDNN model in [51] could not yield results as satisfactory as AVDCNN did in this task. In [51], the testing condition for AVDNN was a noise-matched scenario. Nevertheless, the conditions are much more challenging in this study, and AVDNN appears unable to be as effective as AVDCNN in such a scenario. As shown in Fig. 3 (e) and (g), the ineffective aspects of AVDNN included incompleteness of noise reduction when the mouth was closed, and poor reconstruction of the target speech signals. Such results may stem from the inadequate visual features of AVDNN, suggesting that the visual features learned by CNNs directly from images could be more robust than the handcrafted ones used in AVDNN. To summarize this subsection, the spectrograms in Fig. 3 demonstrate that the proposed AVDCNN model is more powerful than the other baseline SE models, which is also supported by the instrumental measures in the next subsection.

### C. *Results of Instrumental Measures*

In this subsection, we report the results of five SE methods in terms of five instrumental metrics, namely PESQ, STOI, SDI, HASQI, and HASPI. The PESQ measure (ranging from 0.5 to 4.5) indicates the quality measurement of enhanced speech. The STOI measure (ranging from 0 to 1) indicates the intelligibility measurement of enhanced speech. The HASQI and HASPI measures (both ranging from 0 to 1) evaluate sound quality and perception, respectively, for both normal hearing and hearing-impaired people (by setting specific modes). In this study, the normal hearing mode was adopted for both the HASQI and HASPI measures. The SDI measure calculates the distortion measurement of clean and enhanced speech. Except for SDI, larger values indicate a better performance. We report the average evaluation score over the 40 test utterances under different noise types, and SIR and SAR conditions.

We first intended to investigate the SE performances on different noise types. Figs. 4–8 show the average PESQ, STOI, SDI, HASQI, and HASPI scores, respectively, of 10 different SIR noises and the enhanced speech obtained using different SE methods, where the SAR was fixed to 0 dB. From Figs. 4–8, we first notice that the performances of the two conventional SE methods (logMMSE and KLT) show that they cannot effectively handle non-stationary noises. Next, when comparing the two CNN-based models, AVDCNN outperforms ADCNN consistently in terms of all evaluation metrics, confirming the effectiveness of the combination of visual and audio information to achieve a better SE performance. In addition, AVDCNN shows its effectiveness as an audio-visual model by outperforming AVDNN in all the metrics. To further confirm the significance of the superiority of the AVDCNN model over the second best system in each test condition in Figs 4-8, we performed a one-way analysis of variance (ANOVA) and Tukey post-hoc comparisons (TPHCs) [78]. The results confirmed that these scores differed significantly, with p-values of less than 0.05 in most conditions, except for STOI (with music and siren noises), SDI (with baby crying, music, and siren noises), and HASPI (with music and siren noises). With a further analysis on the experimental results, we note that among the 10 testing noise types, the evaluation scores of the baby crying sound are always inferior to those of other noise types, suggesting that the baby crying noise is relatively challenging to handle. Meanwhile, the multiple background talker (2T, 3T) scenarios do not appear to be more challenging than that of the single background talker (1T).

Next, we compared the SE performances provided by different SE models on different SAR levels. Figs. 9–13 show the average PESQ, STOI, SDI, HASQI, and HASPI scores of noisy and the enhanced speech at specific SIR (over 10 different noise types) and SAR levels. In these figures, “×”, “□”, and “○” denote 5 dB, 0 dB, -5 dB SAR, respectively. Please note that a speech signal with a higher SAR indicates that it is involves fewer car engine noise components. From Figs. 9–13, it is clear that (1) the instrumental evaluation results of higher SAR levels are usually better than those of lower SAR levels; and (2) AVDCNN outperforms the other SE methods, which is especially obvious in lower SIR levels. This result shows that visual information provides important clues for assisting SE in AVDCNN in more challenging conditions.

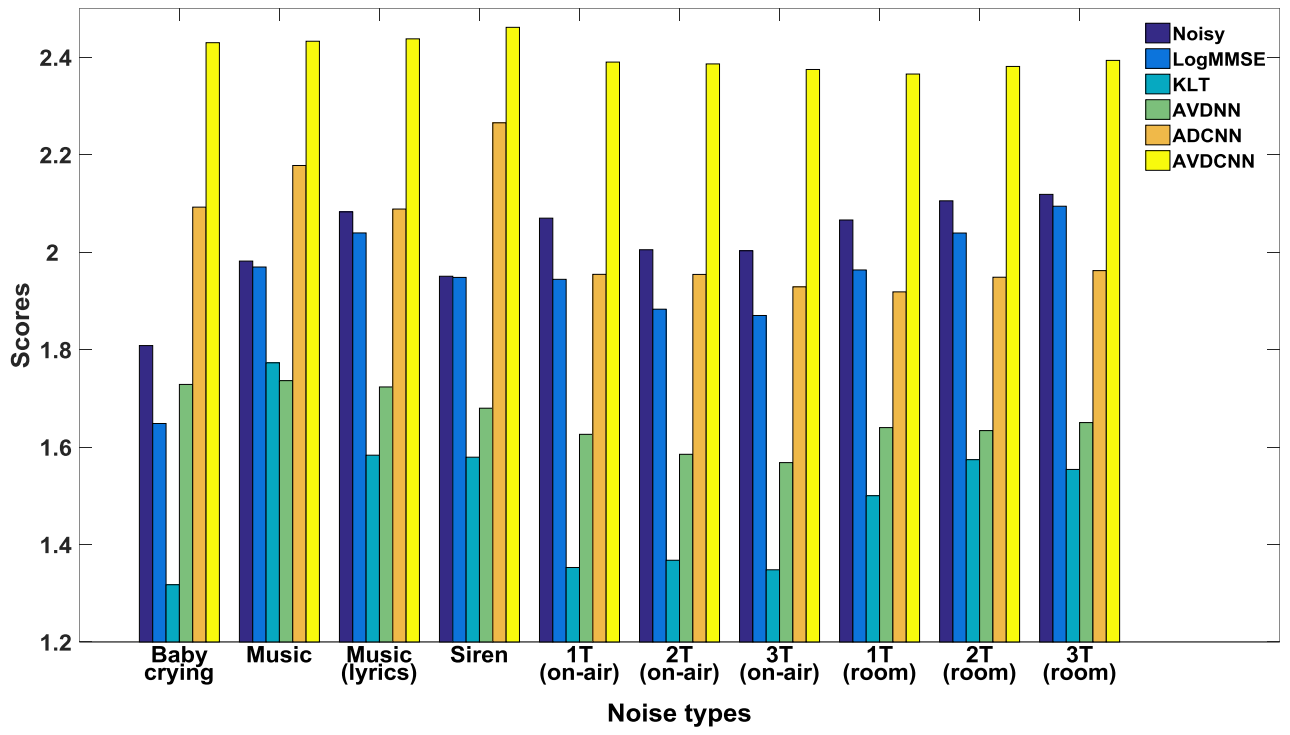

Fig. 4 Mean PESQ scores of 10 different noisy and corresponding enhanced versions of speech, considering different enhancement approaches and varying noise types at an SAR of 0 dB.

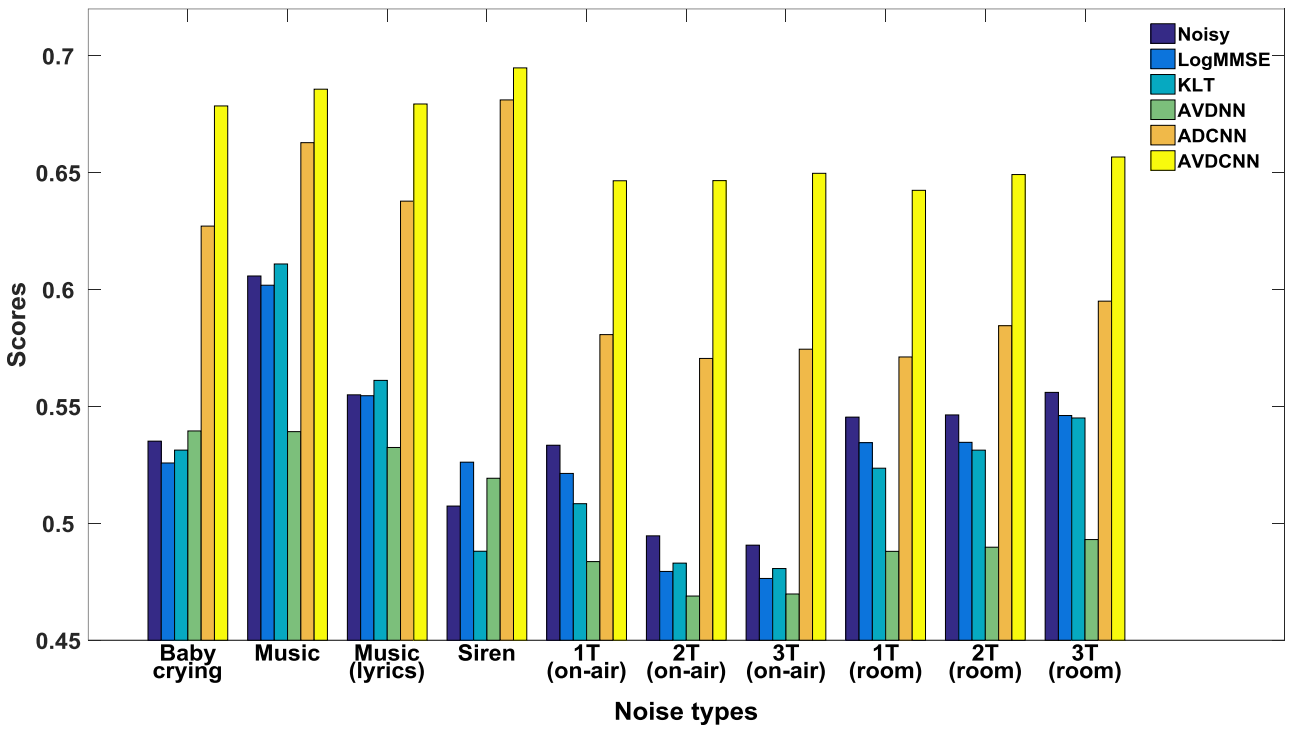

Fig. 5 Mean STOI scores of 10 different noisy and corresponding enhanced versions of speech, considering different enhancement approaches and varying noise types at an SAR of 0 dB.

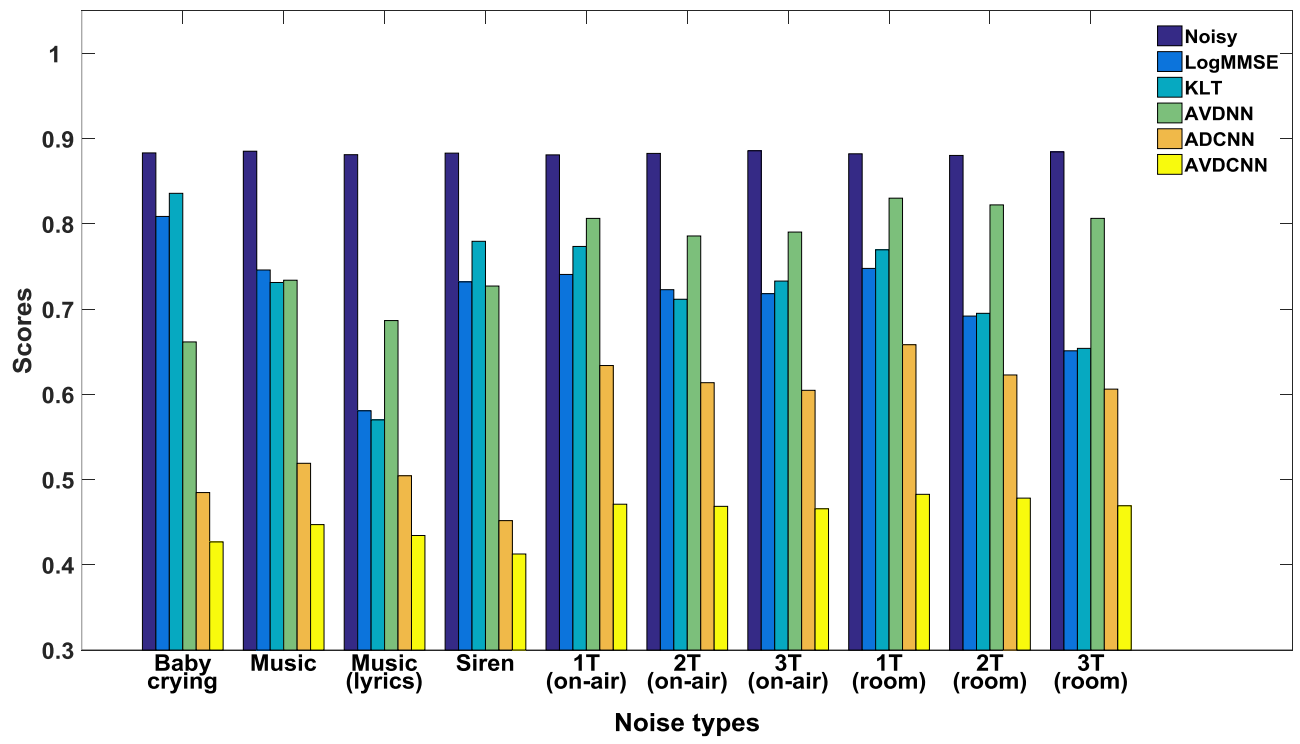

Fig. 6 Mean SDI scores of 10 different noisy and corresponding enhanced versions of speech, considering different enhancement approaches and varying noise types at an SAR of 0 dB.

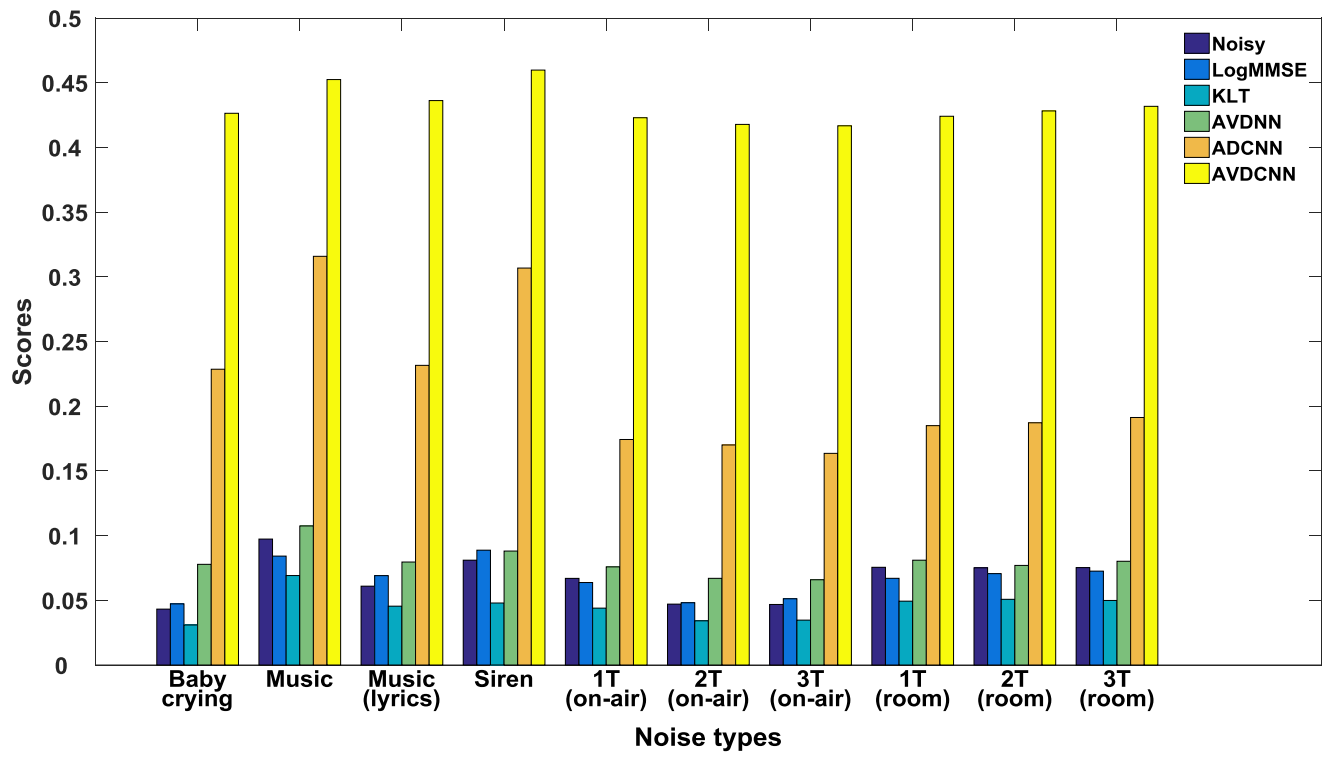

Fig. 7 Mean HASQI scores of 10 different noisy and corresponding enhanced versions of speech, considering different enhancement approaches and varying noise types at an SAR of 0 dB.

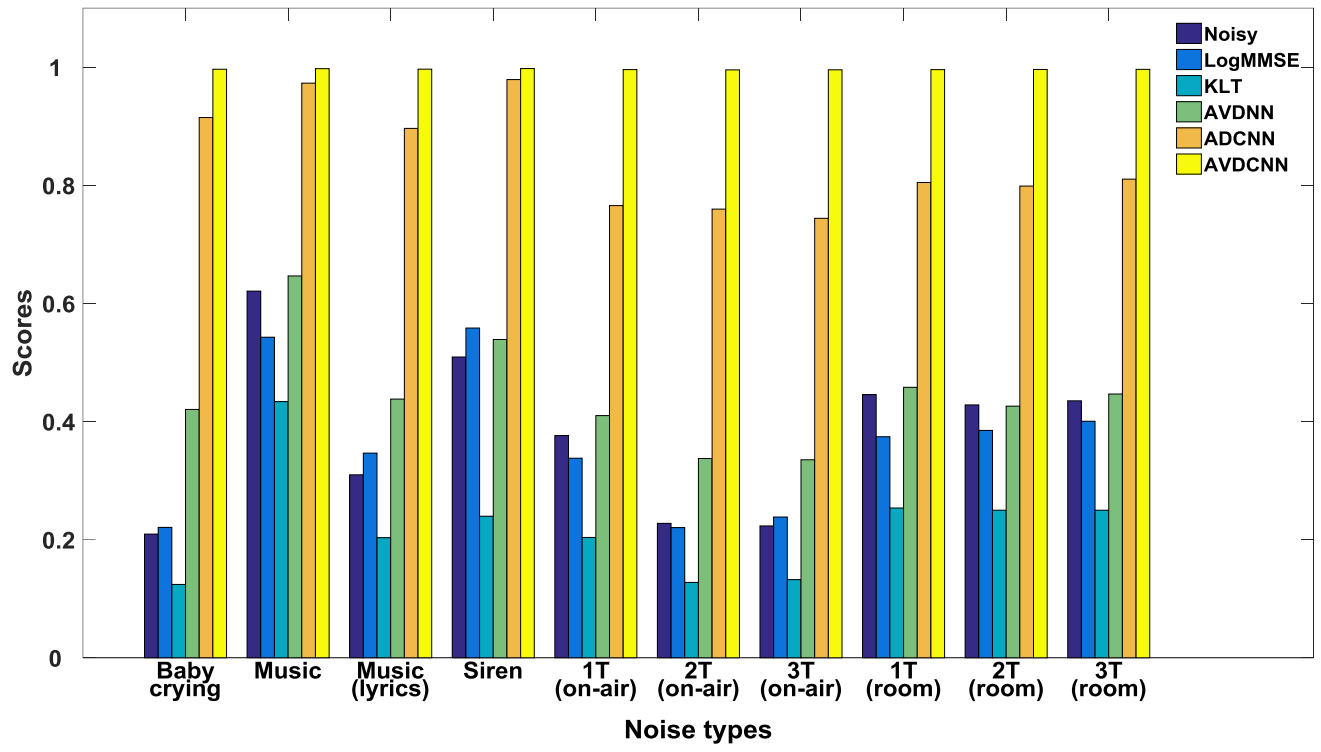

Fig. 8 Mean HASPI scores of 10 different noisy and corresponding enhanced versions of speech, considering different enhancement approaches and varying noise types at an SAR of 0 dB.

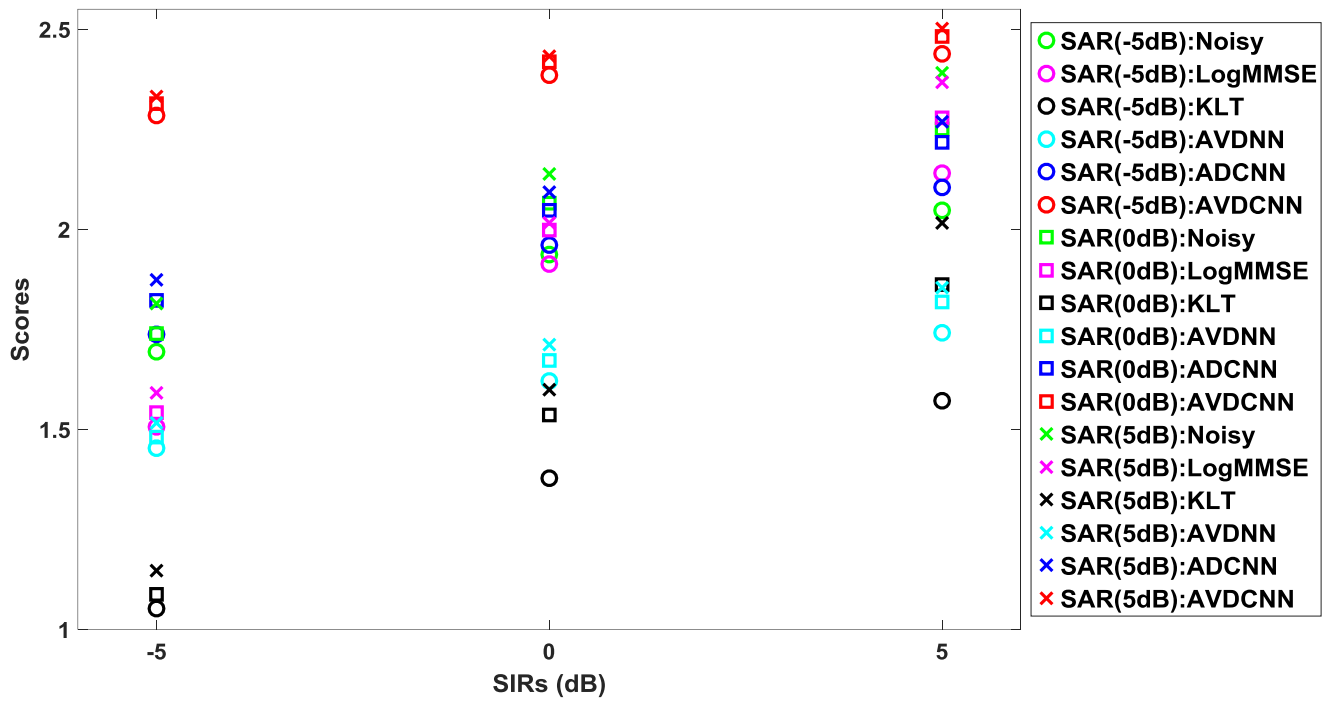

Fig. 9 Mean PESQ scores over 10 different noisy and corresponding enhanced versions of speech, considering different enhancement approaches for each SIR and SAR.

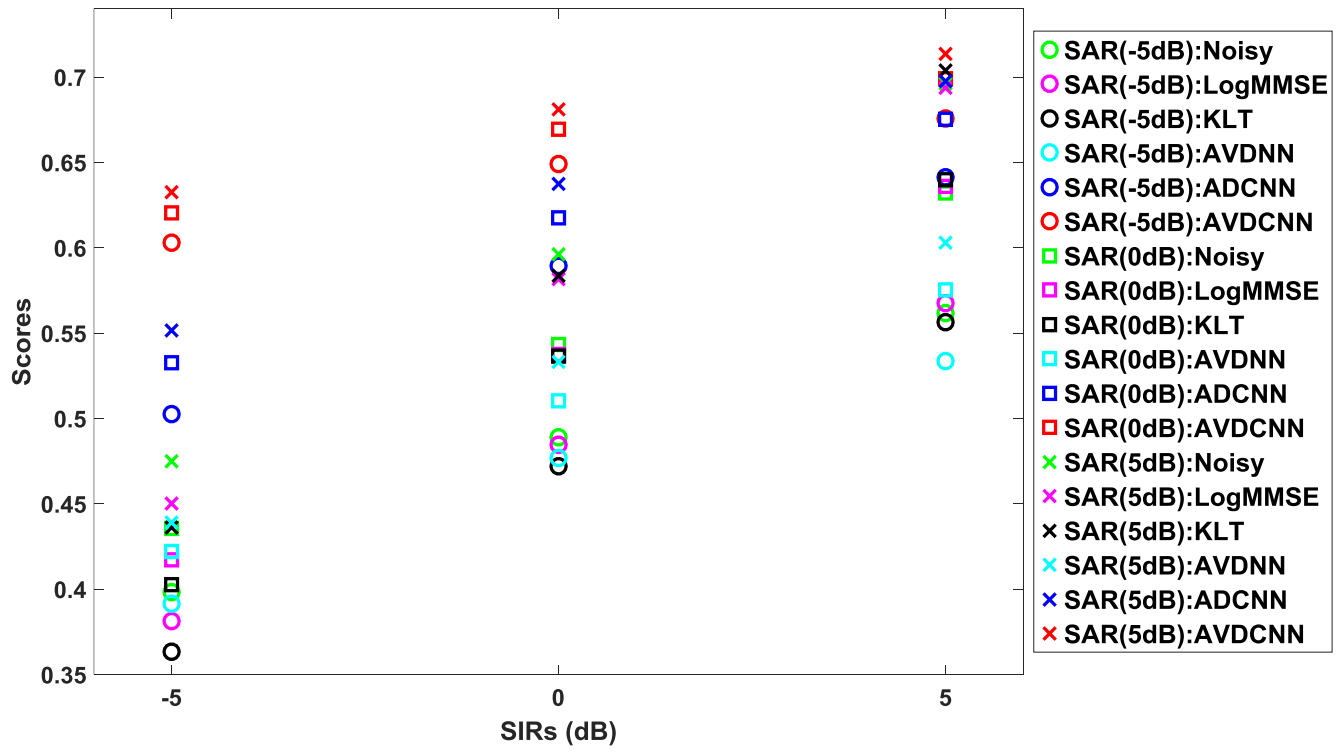

Fig. 10 Mean STOI scores over 10 different noisy and corresponding enhanced versions of speech, considering different enhancement approaches for each SIR and SAR.

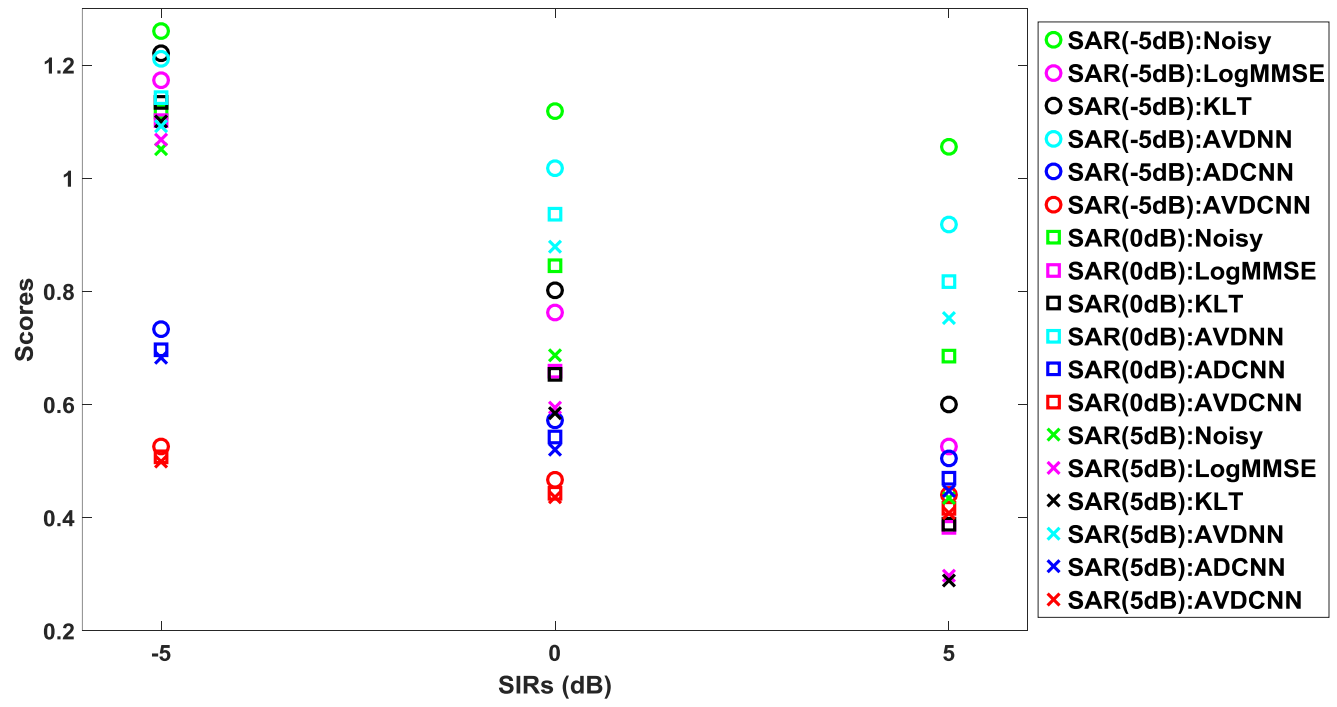

Fig. 11 Mean SDI scores over 10 different noisy and corresponding enhanced versions of speech, considering different enhancement approaches for each SIR and SAR.

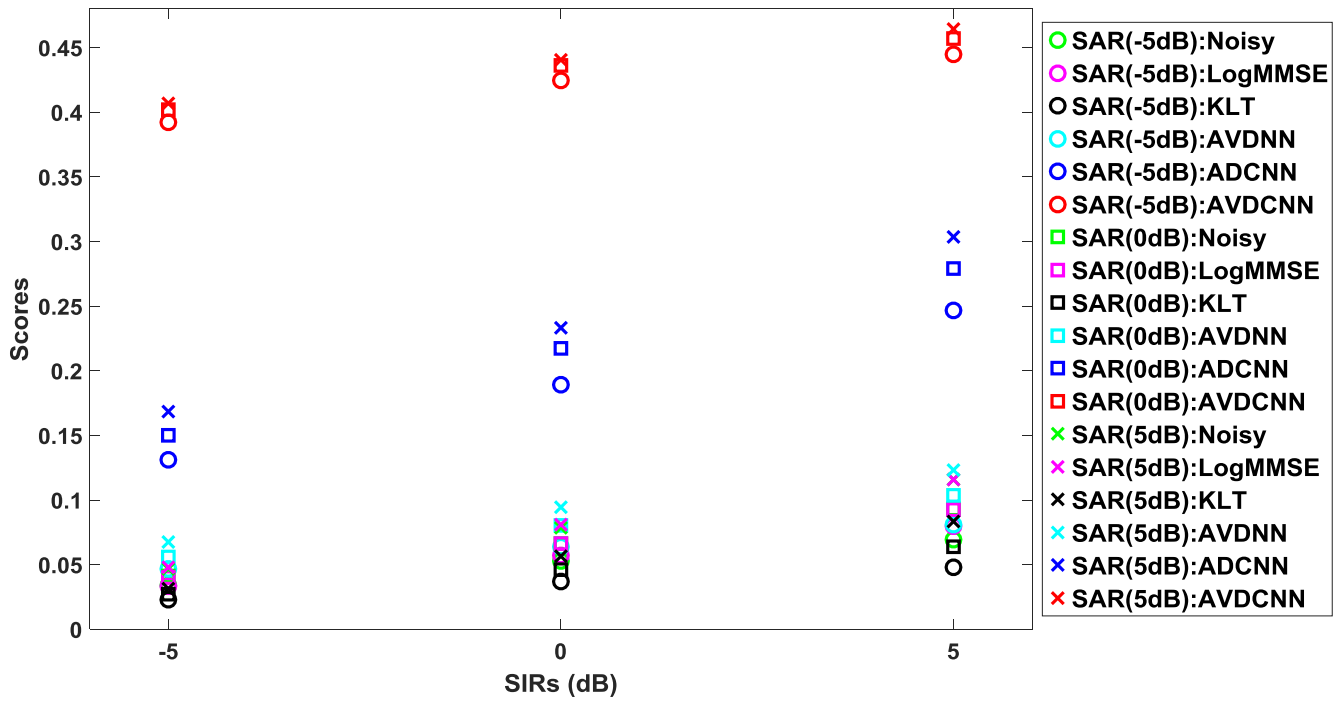

Fig. 12 Mean HASQI scores over 10 different noisy and corresponding enhanced versions of speech, considering different enhancement approaches for each SIR and SAR.

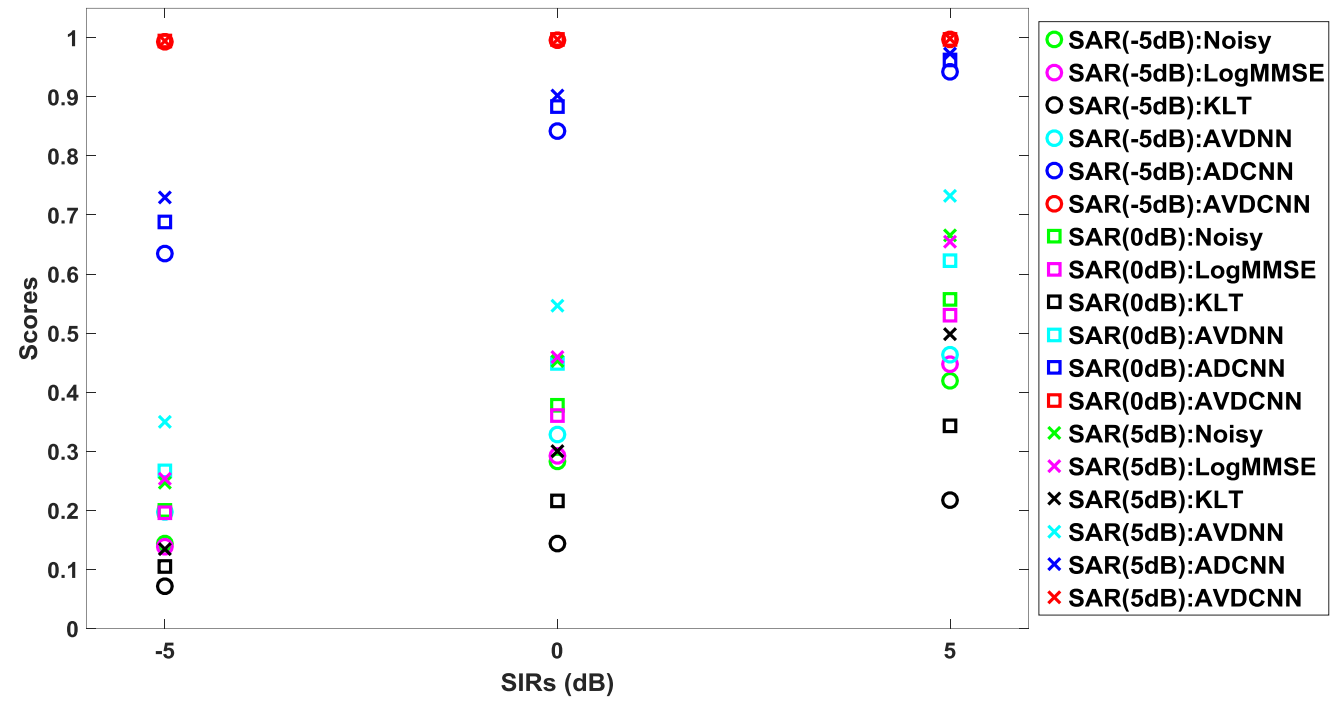

Fig. 13 Mean HASPI scores over 10 different noisy and corresponding enhanced versions of speech, considering different enhancement approaches for each SIR and SAR.

### D. *Multi-style Training Strategy*

A previous study [79] has shown that the input of a certain modality of a multimodal network could dominate over other input types. In our preliminary experiments, we observed similar properties. To alleviate this issue, we adopted the multi-style training strategy [80], which randomly selected the following input types: audio-visual, visual-only, and audio-only, for every 45 epochs in the training phase. When using the visual input only with the audio input set to zeros, a visual output was provided, while audio output was set according to two different models: Model-1 set the audio target to zeros, and Model-2 used the clean audio signals as the target. Similarly, when using the audio-only data with the visual input set to zeros, Model-1 set the visual target to zeros and Model-2 used the original visual data for the visual target. It should be noted that both Model-1 and Model-2 were trained via the multi-style training strategy, and the difference lies in the information specified in the output during the training process. The mean squared errors (MSE) from the training processes of Model-I and Model-II are listed in Figs. 14 and 15, respectively. On the tops of Figs. 14 and 15, we used the bars to mark the epoch segments of the three types of input, namely audio-visual, visual-only, and audio-only.

From the results shown in Figs. 14 and 15, we can observe some support for including visual information. From the windows with solid red line in these two figures, we note that the audio loss was relatively large when we used audio-only data for training. The MSE dropped to a lower level once visual features were used, indicating a strong correlation between audio and visual streams.

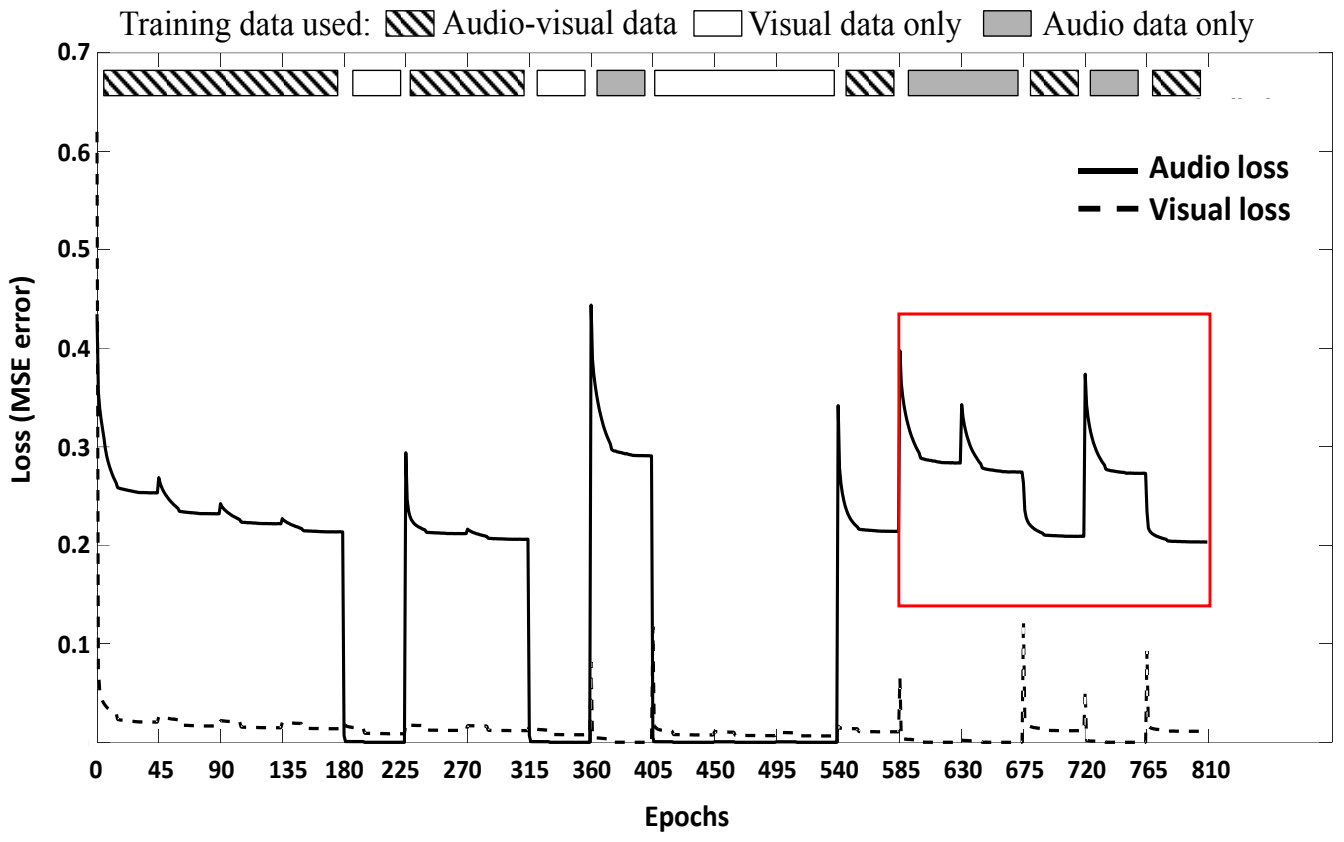

Fig. 14 The learning curve of the training data for the multi-style learning model using Model-I. Model-I means setting the visual/audio target to zeros when only an audio/visual input was selected in training. The red frame shows that a smaller audio loss could be achieved as additional visual information was included.

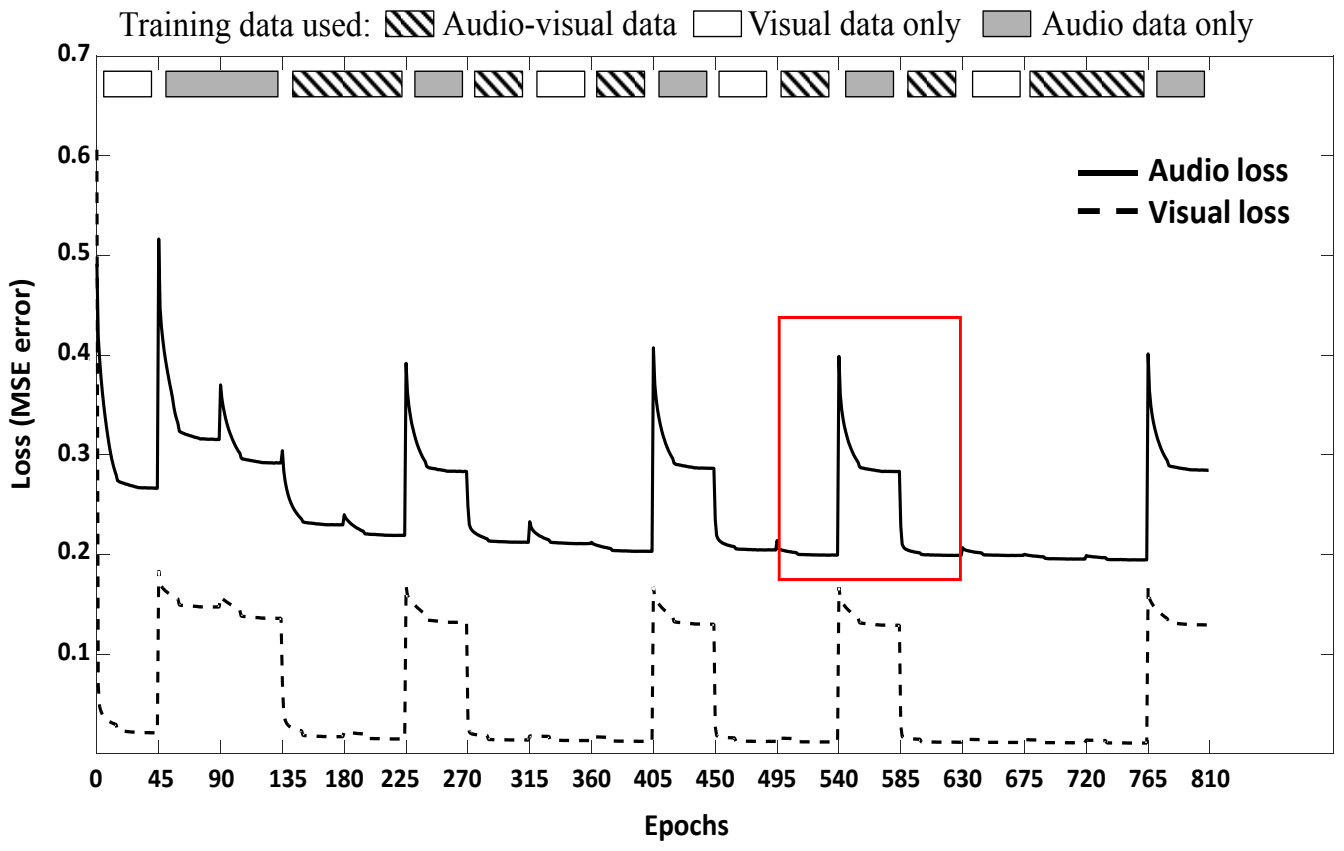

Fig. 15 The learning curve of the training data for the multi-style learning model using Model-II. Model-II means retaining the visual/audio target when only audio/visual input was selected in training. The red frame shows that a smaller audio loss could be achieved as additional visual information was included.

## E. *Mixing Weight*

In the above experiments, the mixing weight in Eq. (5) was fixed to 1. Namely, the errors are considered equally harmful when training the model parameters of AVDCNN. In this subsection, we explore the correlation of $\mu$ with the SE performance. Fig. 16 shows the audio and visual losses in the training data under different mixing weights during the training process of the AVDCNN model. It is observed that the more we emphasized the visual information, i.e., the larger the value of the mixing weight $\mu$, the better visual loss and worse audio loss we obtained. Given that the audio loss dominated the enhancement results, we tended to select a smaller $\mu$.

## F. *Multimodal Inputs with Mismatched Visual Features*

In this subsection, we show the importance of correct matching between input audio features and their visual counterpart features. We selected eight mouth shapes during speech as stationary visual units, and then for each "snapshot" we fixed it as a visual feature for the entire utterance. From the spectrogram in Fig. 17, we can see that the AVDCNN-enhanced speech with correct lip features preserved more detailed structures than other AVDCNN-enhanced speech signals with incorrect lip feature sequences. The mean PESQ score for 40 testing utterances with correct visual features was 2.54, and the mean score of enhanced speech signals with the eight fake lip shape sequences ranged from 1.17 to 2.07. These results suggest that the extraction of the lip shape notably affects the performance of AVDCNN.

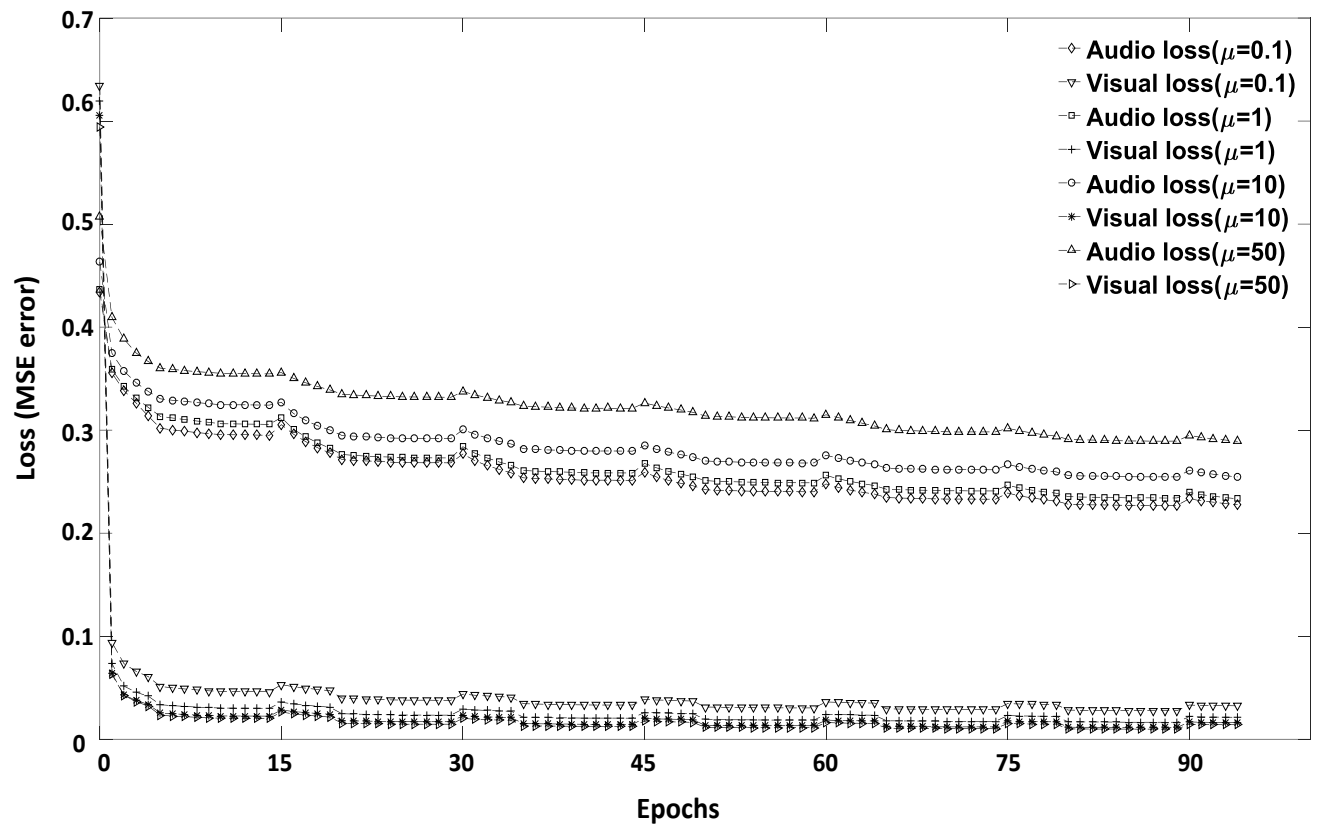

Fig. 16 The audio and visual losses in the training data under different mixing weights during the training process of the AVDCNN model.

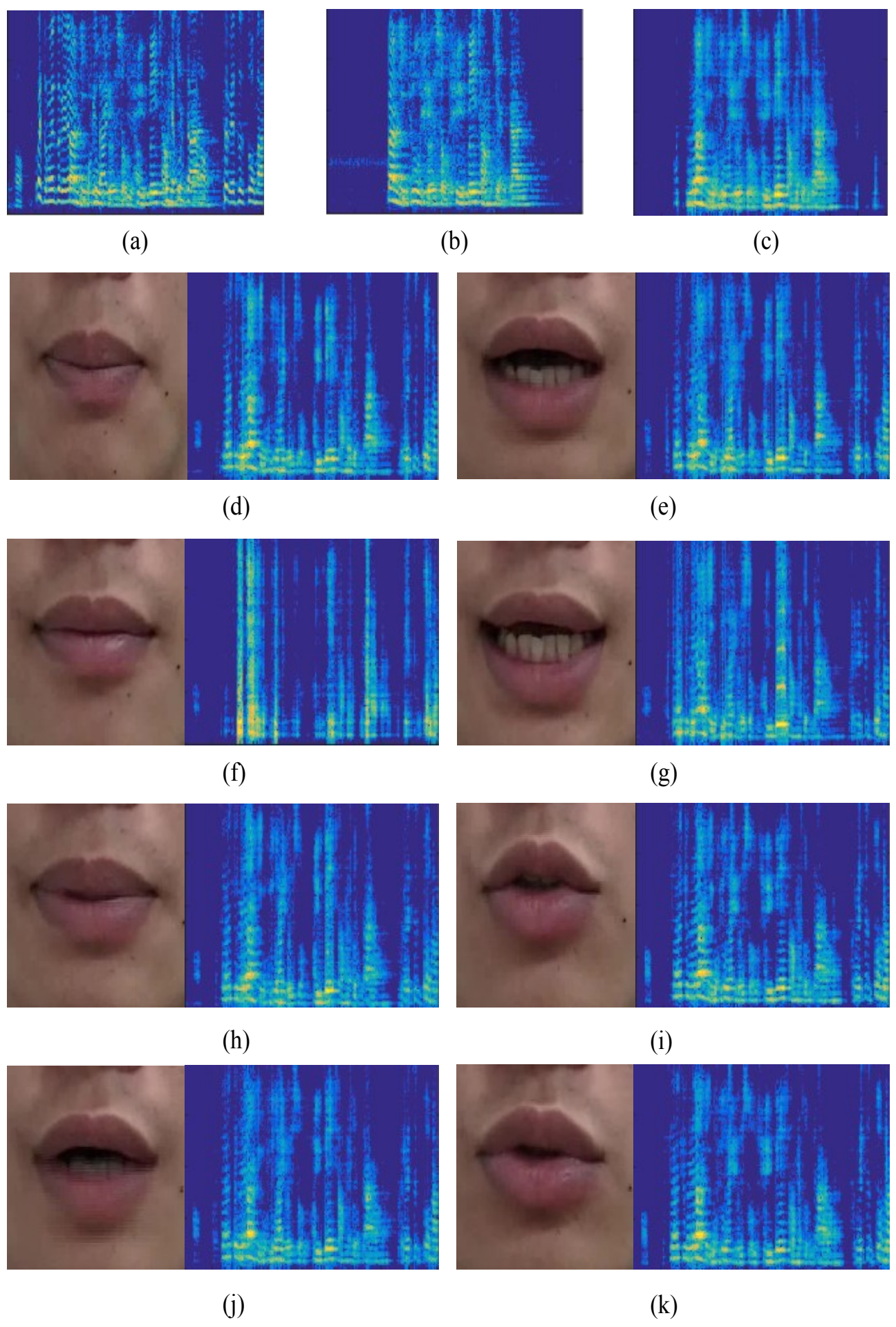

Fig. 17 (a) The noisy speech of 1T (on-air) noise at 0 dB SIR, (b) the clean speech, (c) the AVDCNN-enhanced speech with the correct lip features, (d)-(k) (left) the selected lip shapes, (right) the AVDCNN-enhanced speech with the incorrect lip features, which are sequences of the selected lip shapes.

### G. Reconstructed Mouth Images

In the proposed AVDCNN system, we used visual input as an auxiliary clue for speech signals and added visual information at the output as part of the constraints during the training of the model. Therefore, the proposed system is in fact an audio-visual encoder-decoder system with multi-task learning. In addition to enhanced speech frames, we received the corresponding mouth images at the output in the testing phase. It is interesting to investigate the images obtained using the audio-visual encoder-decoder system. Fig. 18 presents a few visualized samples. For now, we simply view these images as "by-products" of the audio-visual system, compared with the target enhanced speech signals. However, in the future it will be interesting to explore the lip shapes that the model learns when the corresponding fed visual hints are considerably corrupted or not provided.

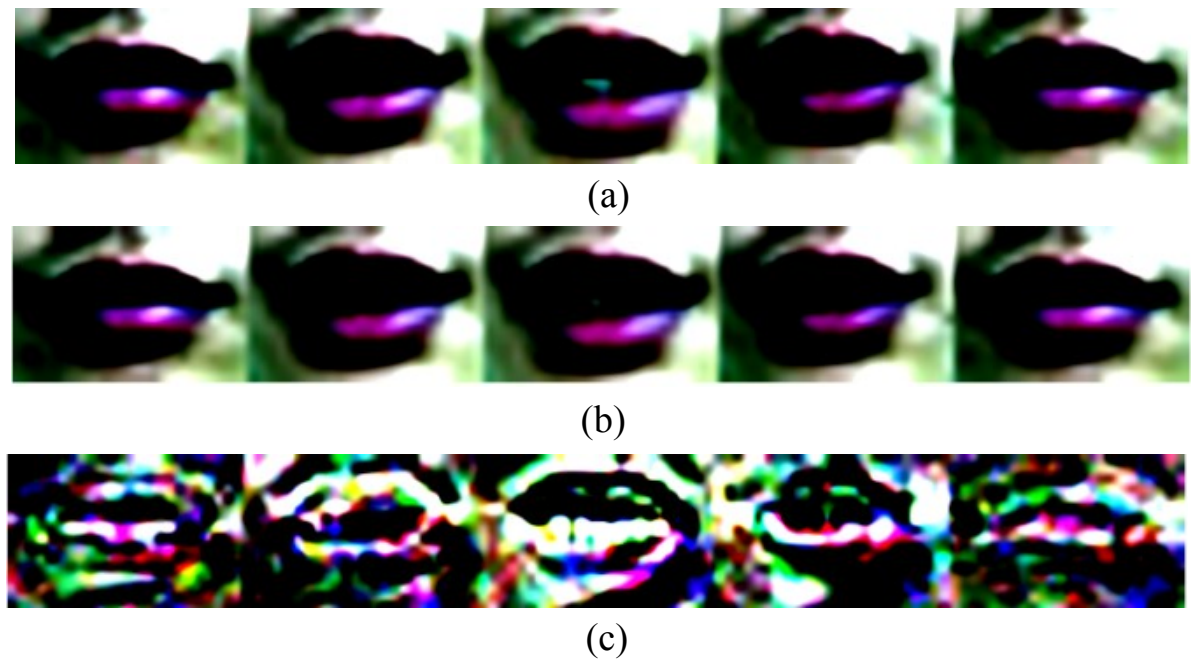

Fig. 18 Visualizing the normalized mouth images. (a) The visual input and (b) visual output of the proposed AVDCNN model. (c) The difference between (a) and (b), with the amplitude magnified ten times.

### H. Subjective Listening Tests

In addition to instrumental evaluations, we also conducted subjective listening tests for enhanced speech from three different methods, namely logMMSE, ADCNN, and AVDCNN. We adopted the procedures for listening tests from [81], using a five-point scale to evaluate the background noise intrusiveness (BAK) and overall effect (OVRL). Higher scores are more favorable. Each subject listened to 10 utterances enhanced from all 10 testing noises under -5 dB SIR and -5 dB SAR by the aforementioned three models, resulting in a total of (3×10×10) utterances. There were a total of 20 subjects, whose native language is Mandarin, participating in the tests. The subjects were between 23 and 40 years old, with a mean of 26 years. The mean scores over the subjects are presented in Table II. These results show that the proposed AVDCNN model obtained the best scores among the three models compared in the subjective listening tests.

TABLE II
RESULTS OF THE SUBJECTIVE LISTENING TESTS COMPARING THE THREE DIFFERENT SE MODELS

| Models | BAK | OVRL |
|---|---|---|
| LogMMSE | 1.20 | 1.70 |
| ADCNN | 2.75 | 1.95 |
| AVDCNN | **3.70** | **2.95** |

### I. Early Fusion Scheme for the AVDCNN Model

We also attempted an early fusion scheme, by combining audio and visual features at inputs before entering the convolutional layers. The early fusion model, denoted by AVDCNN-EF, replaced the audio network and visual network in Fig. 1 with united CNNs, whose input consisted of the fused audio-visual features generated by concatenating audio features, separated RGB channels of visual features, and zero paddings, with a final shape of 257×29×1 (audio:257×5×1, RGB:(80+80+80)×24×1, and zero padding:17×24×1). The numbers of parameters of AVDCNN and AVDCNN-EF are of the same order. A comparison of the instrumental metrics of the enhanced results for AVDCNN and AVDCNN-EF is presented in Table III. The scores represent the mean scores for the enhanced speech over 10 different noises under different SIRs at an SAR of 0 dB. It is clear that AVDCNN consistently outperforms AVDCNN-EF, indicating that the proposed fusion scheme, which processes audio and visual information individually first and fuses them later, is better than an early fusion scheme, which combines the heterogeneous data at the beginning.

TABLE III
MEAN SCORES OF THE INSTRUMENTAL METRICS OF THE ENHANCED SPEECH OVER 10 DIFFERENT NOISES UNDER DIFFERENT SIRS AT 0 DB SAR, COMPARING THE AVDCNN MODELS WITH AND WITHOUT EARLY FUSION

| Models | PESQ | STOI | SDI | HASQI | HASPI |
|---|---|---|---|---|---|
| AVDCNN | **2.41** | **0.66** | **0.45** | **0.43** | **0.99** |
| AVDCNN-EF | 1.52 | 0.51 | 1.43 | 0.11 | 0.74 |

## V. DISCUSSION

From the previous experiments, we can observe clear evidence of how visual information can affect the enhancement results. For instance, Fig. 3, (g) shows that noise and speech signals from a non-targeted speaker were effectively suppressed when the mouth was closed. This result indicates that visual information plays a beneficial role in voice activity detection (VAD). In fact, there are researchers working in this particular direction [82, 83]. This also contributes to why we choose in-car environments as our testing scenario, and investigate the effectiveness of audio-visual SE. If there is a camera that targets a driver's mouth region, the lip shape could provide a strong hint on whether or not to activate a voice command system with background talkers or noises, and in addition could enhance the speech. The lip shape could provide a useful hint for VAD. However, this does not yet appear to be a very solid one. As shown in Fig. 19, in a few of the testing results for enhanced speech using the AVDCNN model we observed that noise components were incompletely removed in the non-speech segment, because of the open shape of the mouth at that time. We believe that this shortcoming could be further improved with the combination of audio-only VAD techniques.

We also preliminarily evaluated the AVDCNN model on real-world testing data, i.e., the noisy speech was recorded in a real noisy environment, rather than artificially adding noises to the clean speech. Fig. 20 (a) illustrates the controlled environment for recording training and testing data. Fig. 20 (b) illustrates the recording conditions of the real-world data, which was recorded by a smart phone (ASUS ZenFone 2 ZE551ML) in a night market. The spectrograms of the noisy and AVDCNN-enhanced speech signals are presented in Fig. 20 (c) and (d), respectively. The red frame indicates the segment where the target speech was mixed with the background talking sound. We observe that the lip shape helped to identify the target speech segment, while the reconstruction on the target speech was not as good as the enhanced results in the controlled environment. This might be because of different light coverage, a lower SIR, or properties of the background noise, suggesting that there remains room for improvement in audio-visual SE in real-world testing conditions.

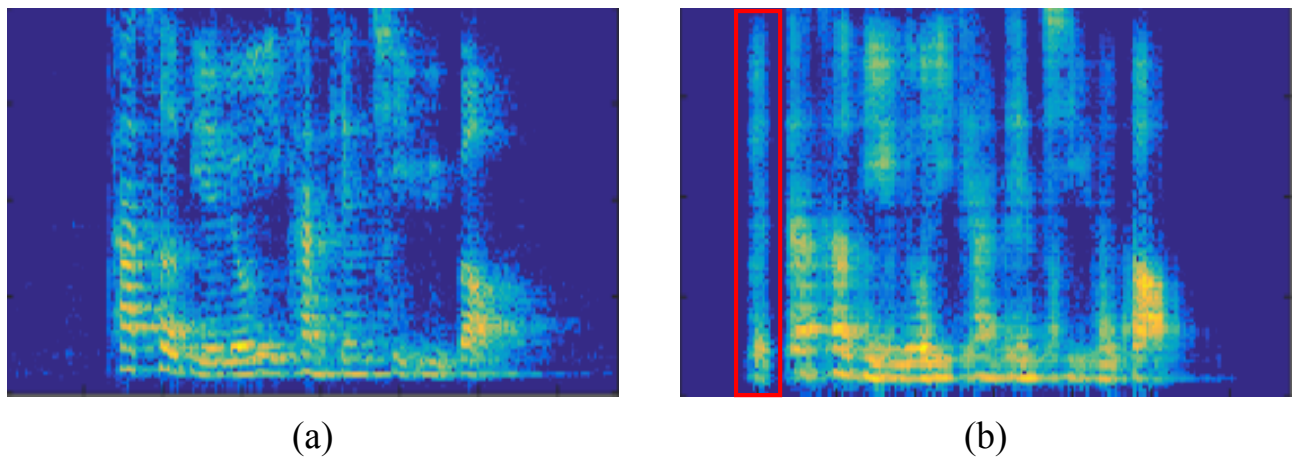

Fig. 19 Spectrograms of (a) the clean speech and (b) the AVDCNN-enhanced speech. The red frame in (b) shows that noise was reduced incompletely in the non-speech segment if the mouth was in an unclosed shape.

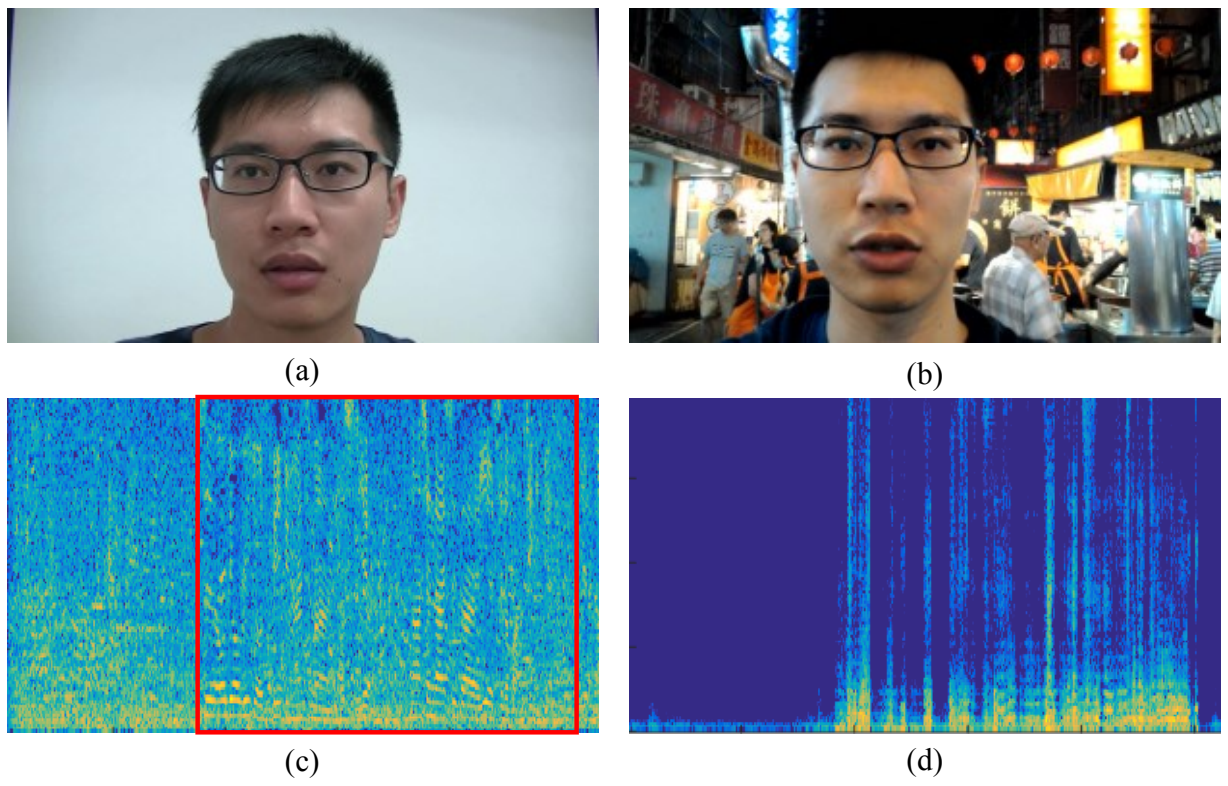

Fig. 20 Testing in the real-world conditions. (a) The controlled environment (seminar room) for recording the training and testing data. (b) The recording environment (night market) for the real-world test data. Spectrograms of (c) the noisy speech with the babble noise and (d) the enhanced speech from the AVDCNN model. The red frame in (c) indicates the segment where the target speech was mixed with noise.

## VI. Conclusion

In this paper, we have proposed a novel CNN-based audio-visual encoder-decoder system with multi-task learning for speech enhancement, called AVDCNN. The model utilizes individual networks to process input data with different modalities, and a fusion network is then employed to learn joint multimodal features. We trained the model in an end-to-end manner. The experimental results obtained using the proposed architecture show that its performance for the SE task is superior to that of three audio-only baseline models in terms of five instrumental evaluation metrics, confirming the effectiveness of integrating visual information with audio information into the SE process. We also demonstrated the model's effectiveness by comparing it with other audio-visual SE models. Overall, the contributions of this paper are five-fold. First, we adopted CNNs for audio and visual streams in the proposed end-to-end audio-visual SE model, obtaining improvements over many baseline models. Second, we quantified the advantages of integrating visual information for SE through the multi-modal and multi-task training strategies. Third, we demonstrated that processing audio and visual streams with late fusion is better than early fusion. Fourth, the experimental results exhibited a high correlation between speech and lip shape, and showed the importance of using correct lip shapes in audio-visual SE. Finally, we showed that lip shapes were effective as auxiliary features in VAD, and also pointed out the potential problems when using audio-visual SE models. In the future, we will attempt to improve the proposed architecture by using a whole face as visual input, rather than the mouth region only, in order to exploit well-trained face recognition networks to improve visual descriptor networks. Furthermore, we plan to modify the existing CNNs in our model by considering other state-of-the-art CNN-based models, such as fully convolutional networks [84-86] and U-Net [87]. A more sophisticated method for synchronizing the audio and video streams might improve the performance and is worthy of investigation. Finally, to improve the practicality of the model for real-world application scenarios, we will consider collecting training data including more complicated and real conditions.

## Acknowledgment

This work was supported by the Academia Sinica Thematic Research Program AS-105-TP-C02-1.

## References

[1] J. Li, L. Deng, R. Haeb-Umbach, and Y. Gong, *Robust Automatic Speech Recognition: A Bridge to Practical Applications*, 1st ed. Academic Press, 2015.

[2] B. Li, Y. Tsao, and K. C. Sim, "An investigation of spectral restoration algorithms for deep neural networks based noise robust speech recognition," in *Proc. INTERSPEECH*, 2013, pp. 3002–3006.

[3] A. El-Solh, A. Cuhadar, and R. A. Goubran. "Evaluation of speech enhancement techniques for speaker identification in noisy environments," in *Proc. ISMW*, 2007, pp. 235–239.

[4] J. Ortega-Garcia and J. Gonzalez-Rodriguez, "Overview of speech enhancement techniques for automatic speaker recognition," in *Proc. ICSLP*, vol. 2, 1996, pp. 929–932.

[5] J. Li, L. Yang, Y. Hu, M. Akagi, P.C. Loizou, J. Zhang, and Y. Yan, "Comparative intelligibility investigation of single-channel noise reduction algorithms for Chinese, Japanese and English," *Journal of the Acoustical Society of America*, 2011, vol. 129, no. 5, pp. 3291–3301.

[6] J. Li, S. Sakamoto, S. Hongo, M. Akagi, and Y. Suzuki, "Two-stage binaural speech enhancement with Wiener filter for high-quality speech communication," *Speech Communication*, vol. 53, no. 5, pp. 677–689, 2011.

[7] T. Venema, *Compression for Clinicians*, 2nd ed. Thomson Delmar Learning, 2006, chapter. 7.

[8] H. Levitt, "Noise reduction in hearing aids: an overview," *J. Rehab. Res. Dev.*, 2001, vol. 38, no. 1, pp.111–121.

[9] A. Chern, Y. H. Lai, Y.-P. Chang, Yu Tsao, R. Y. Chang, and H.-W. Chang, "A smartphone-based multi-functional hearing assistive system to facilitate speech recognition in the classroom," *IEEE Access*, 2017.

[10] Y. H. Lai, F. Chen, S.-S. Wang, X. Lu, Y. Tsao, and C.-H. Lee, "A deep denoising autoencoder approach to improving the intelligibility of vocoded speech in cochlear implant simulation," *IEEE Transactions on Biomedical Engineering*, vol.64, no. 7, pp. 1568–1578, 2016.

[11] F. Chen, Y. Hu, and M. Yuan, "Evaluation of noise reduction methods for sentence recognition by Mandarin-speaking cochlear implant listeners," *Ear and Hearing*, vol. 36, no.1, pp. 61–71, 2015.

[12] Y. H. Lai, Y. Tsao, X. Lu, F. Chen, Y.-T. Su, K.-C. Chen, Y.-H. Chen, L.-C. Chen, P.-H. Li, and C.-H. Lee, "Deep learning based noise reduction approach to improve speech intelligibility for cochlear implant recipients," to appear in *Ear and Hearing*.

[13] J. Chen, "Fundamentals of Noise Reduction," in *Spring Handbook of Speech Processing*, Springer, 2008, chapter. 43.

[14] P. Scalart and J. V. Filho, "Speech enhancement based on a priori signal to noise estimation," in *Proc. ICASSP*, 1996, pp. 629–632.

[15] Y. Ephraim and D. Malah, “Speech enhancement using a minimum mean-square error short-time spectral amplitude estimator,” *IEEE Transactions on Acoustics, Speech and Signal Processing*, vol. 32, no. 6, pp. 1109–1121, 1984.

[16] R. Martin, “Speech enhancement based on minimum mean-square error estimation and supergaussian priors,” *IEEE Transactions on Speech and Audio Processing*, vol. 13, no. 5, pp. 845–856, 2005.

[17] Y. Tsao and Y. H. Lai, “Generalized maximum a posteriori spectral amplitude estimation for speech enhancement,” *Speech Communication*, vol. 76, pp. 112–126, 2015.

[18] A. Hussain, M. Chetouani, S. Squartini, A. Bastari, and F. Piazza, “Nonlinear speech enhancement: An overview,” in *Progress in Nonlinear Speech Processing. Berlin*, Germany: Springer, 2007, pp. 217–248.

[19] A Uncini., “Audio signal processing by neural networks”, *Neurocomputing*, vol. 55, pp. 593–625, 2003.

[20] G. Cocchi and A. Uncini, “Subband neural networks prediction for on-line audio signal recovery,” *IEEE Transactions on Neural Networks*, vol. 13, no. 4, pp. 867–876, 2002.

[21] N. B. Yoma, F. McInnes, M. Jack, “Lateral inhibition net and weighted matching algorithms for speech recognition in noise,” *Proc. IEE Vision, Image & Signal Processing*, vol. 143, no. 5, pp. 324–330, 1996.

[22] X. Lu, Y. Tsao, S. Matsuda, and C. Hori, “Speech enhancement based on deep denoising autoencoder,” in *Proc. INTERSPEECH*, 2013, pp. 436–440.

[23] X. Lu, Y. Tsao, S. Matsuda, and C. Hori, “Ensemble modeling of denoising autoencoder for speech spectrum restoration,” in *Proc. INTERSPEECH*, 2014, pp. 885–889.

[24] Y. Xu, J. Du, L. R. Dai, and C. H. Lee, “An experimental study on speech enhancement based on deep neural networks,” *IEEE Signal Processing Letters*, 2014, vol. 21, pp. 65–68.

[25] D. Liu, P. Smaragdis, and M. Kim, “Experiments on deep learning for speech denoising,” in *Proc. INTERSPEECH*, 2014, pp. 2685–2689.

[26] M. Kolbæk, Z.-H. Tan, and J. Jensen, “Speech intelligibility potential of general and specialized deep neural network based speech enhancement systems,” *IEEE/ACM Transactions on Audio, Speech, and Language Processing*, vol. 25, pp. 153-167, 2017.

[27] F. Weninger, F. Eyben, and B. Schuller, “Single-channel speech separation with memory-enhanced recurrent neural networks,” in *Proc. ICASSP*, 2014, pp. 3709–3713.

[28] F. Weninger, H. Erdogan, S. Watanabe, E. Vincent, J. L. Roux, J. R. Hershey, and B. Schuller, “Speech enhancement with LSTM recurrent neural networks and its application to noise-robust ASR,” in *Latent Variable Analysis and Signal Separation*, pp. 91–99. Springer, 2015.

[29] P. Campolucci, A. Uncini, F. Piazza, and B. Rao, ‘‘On-line learning algorithms for locally recurrent neural networks,’’ *IEEE Transactions on Neural Networks*, vol. 10, no. 2, pp. 253–271, 1999.

[30] F. Eyben, F. Weninger, S. Squartini, and B. Schuller, “Real-life voice activity detection with LSTM recurrent neural networks and an application to Hollywood movies,” in *Proc. ICASSP*, 2013, pp. 483–487.

[31] S.-W. Fu, Y. Tsao, and X. Lu, “SNR-Aware convolutional neural network modeling for speech enhancement,” in *Proc. INTERSPEECH*, 2016.

[32] S.-W. Fu, Y. Tsao, and X. Lu, “Complex spectrogram enhancement by convolutional neural network with multi-metrics learning,” in *Proc. MLSP, 2017*.

[33] H. McGurk and J. MacDonald, “Hearing lips and seeing voices,” *Nature*, vol. 264, pp. 746–748, 1976.

[34] D. G. Stork and M. E. Hennecke, *Speechreading by Humans and Machines*, Springer, 1996.

[35] G. Potamianos, C. Neti, G. Gravier, A. Garg, and Andrew W, “Recent advances in the automatic recognition of audio-visual speech,” *Proceedings of IEEE*, vol. 91, no. 9, 2003.

[36] D. Kolossa, S. Zeiler, A. Vorwerk, and R. Orglmeister, “Audiovisual speech recognition with missing or unreliable data,” in *Proc. AVSP*, 2009, pp. 117–122.

[37] A. V. Nefian, L. Liang, X. Pi, X. Liu, and K. Murphy, “Dynamic Bayesian networks for audio-visual speech recognition,” *EURASIP Journal on Applied Signal Processing*, vol. 2002, no. 11, pp.1274–1288, 2002.

[38] A. H. Abdelaziz, S. Zeiler, and D. Kolossa, “Twin-HMM-based audio-visual speech enhancement,” in *Proc. ICASSP*, 2013, pp. 3726–3730.

[39] S. Deligne, G. Potamianos, and C. Neti, “Audio-visual speech enhancement with AVCDCN (audio-visual codebook dependent cepstral normalization),” in *Proc. Int. Conf. Spoken Lang. Processing,* 2002, pp. 1449–1452.

[40] H. Meutzner, N. Ma, R. Nickel, C. Schymura, and D. Kolossa, “Improving audio-visual speech recognition using deep neural networks with dynamic stream reliability estimates,” in *Proc. ICASSP*, 2017.

[41] V. Estellers, M. Gurban, and J.-P. Thiran, “On dynamic stream weighting for audio-visual speech recognition,” *IEEE Transactions on Audio, Speech, and Language Processing*, vol. 20, no. 4, pp. 1145–1157, 2012.

[42] J. Ngiam, A. Khosla, M. Kim, J. Nam, H. Lee, and A. Ng, “Multimodal deep learning,” in *Proc. ICML*, 2011.

[43] Y. Mroueh, E. Marcheret, and V. Goel, “Deep multimodal learning for audio-visual speech recognition,” in *Proc. ICASSP*, 2015.

[44] L. Girin, J.-L. Schwartz, and G. Feng, “Audio-visual enhancement of speech in noise,” *Journal of the Acoustical Society of America*, vol. 109, pp. 3007, 2001.

[45] R. Goecke, G. Potamianos, and C. Neti, “Noisy audio feature enhancement using audio-visual speech data,” in *Proc. ICASSP*, 2002.

[46] I. Almajai and B. Milner, “Enhancing audio speech using visual speech features,” in *Proc. INTERSPEECH*, 2009.

[47] I. Almajai and B. Milner, “Visually derived Wiener filters for speech enhancement,” *IEEE Transactions on Audio, Speech, and Language Processing*, vol. 19, no.6, pp. 1642–1651, 2011.

[48] B. Rivet, L. Girin, and C. Jutten, “Visual voice activity detection as a help for speech source separation from convolutive mixtures,” *Speech Communication*, vol. 49, no. 7-8, pp. 667–677, 2007.

[49] B. Rivet, L. Girin, and C. Jutten, "Mixing audiovisual speech processing and blind source separation for the extraction of speech signals from convolutive mixtures." *IEEE Transactions on Audio, Speech, and Language Processing* , vol. 15, no. 1, pp. 96–108, 2007.

[50] B. Rivet, W. Wang, S. M. Naqvi, and J. A. Chambers, “Audiovisual speech source separation: An overview of key methodologies,” *IEEE Signal Processing Magazine*, vol. 31, no. 3, pp. 125–134, May 2014.

[51] J.-C. Hou, S.-S. Wang, Y. H. Lai, J.-C. Lin, Y. Tsao, H.-W. Chang, and H.-M. Wang, “Audio-visual speech enhancement using deep neural networks,” in *Proc. APSIPA ASC*, 2016.

[52] Z. Wu, S. Sivadas, Y. K. Tan, B. Ma, and S. M. Goh, “MultiModal hybrid deep neural network for speech enhancement,” arXiv:1606.04750, 2016.

[53] G. Tzimiropoulos and M. Pantic, “Gauss-Newton deformable part models for face alignment in-the-wild,” in *Proc. CVPR*, 2014, pp. 1851–1858.

[54] K. Simonyan and A. Zisserman, “Very deep convolutional networks for large-scale image recognition,” in *Proc. ICLR*, 2015.

[55] O. M. Parkhi, A. Vedaldi, and A. Zisserman, “Deep face recognition,” in *Proc. BMVC*, 2015.

[56] A. W. Rix, J. G. Beerends, M. P. Hollier, and A. P. Hekstra, “Perceptual evaluation of speech quality (PESQ) – a new method for speech quality assessment of telephone networks and codecs,” in *Proc. ICASSP*, 2001.

[57] C. Taal, R. Hendriks, R. Heusdens, and J. Jensen, “An algorithm for intelligibility prediction of time–frequency weighted noisy speech,” *IEEE Transactions on Acoustics, Speech and Signal Processing*, vol. 19, pp. 2125–2136, 2011.

[58] J. Chen, J. Benesty, Y. Huang, and S. Doclo, “New insights into the noise reduction Wiener filter,” *IEEE/ACM Transactions on Audio, Speech, and Language Processing*, vol. 14, pp. 1218–1234, 2006.

[59] J. M. Kates and K. H. Arehart, “The hearing-aid speech quality index (HASQI),” *Journal of the Audio Engineering Society*, vol. 58, no. 5, pp. 363–381, 2010.

[60] J. M. Kates and K. H. Arehart, “The hearing-aid speech perception index (HASPI),” *Speech Communication*, vol. 65, pp. 75–93, 2014.

[61] M. W. Huang, “Development of Taiwan Mandarin hearing in noise test,” Master thesis, Department of speech language pathology and audiology, National Taipei University of Nursing and Health science, 2005.

[62] P. Viola and M. J. Jones, “Robust Real-Time Face Detection,” *International Journal of Computer Vision*, vol. 57, no. 2, pp. 137–154, 2004.

[63] R. Caruana, "Multitask learning," *Machine learning,* vol. 28, pp. 41–75, 1997.

[64] M. L. Seltzer and J. Droppo, “Multi-task learning in deep neural networks for improved phoneme recognition,” in *Proc. ICASSP*, 2013, pp. 6965–6969.

[65] A. Rezayee, and S. Gazor, “An adaptive KLT approach for speech enhancement,” *IEEE Transactions on Speech and Audio Processing*, vol. 9, no. 2, pp. 87–95, 2001.

[66] Y. Ephraim, and D. Malah, “Speech enhancement using a minimum mean-square error log-spectral amplitude estimator,” *IEEE Transactions on Acoustics, Speech and Signal Processing*, vol. 33, no. 2, pp.443–445, 1985.
[67] E. Principi, S. Cifani, R. Rotili, S. Squartini, F. Piazza, “Comparative evaluation of single-channel mmse-based noise reduction schemes for speech recognition,” *Journal of Electrical and Computer Engineering,* pp. 1–7, 2010.
[68] J. Hong, S. Park, S. Jeong, and M. Hahn, “Dual-microphone noise reduction in car environments with determinant analysis of input correlation matrix,” *IEEE Sensors Journal*, vol. 16, no. 9, pp.3131–3140, 2016.
[69] Y. Xu, J. Du, L. R. Dai, and C.-H. Lee, “A regression approach to speech enhancement based on deep neural networks,” *IEEE/ACM Transactions on Audio, Speech and Language Processing*, vol. 23, no. 1, pp. 7–19, 2015.
[70] G. Hu, 100 nonspeech environmental sounds, 2004 [Online]. Available: http://web.cse.ohio-state.edu/pnl/corpus/HuNonspeech/HuCorpus.html.
[71] B. Lee, M. Hasegawa-Johnson, C. Goudeseune, S. Kamdar, S. Borys, M. Liu, and T. Huang, “AVICAR: Audio-visual speech corpus in a car environment,” in *Proc. Int. Conf. Spoken Language*, 2004, pp.2489–2492.
[72] R. Navarathna, D. Dean, S. Sridharan, and P. Lucey, “Multiple cameras for audio-visual speech recognition in an automotive environment,” *Computer Speech & Language*, vol. 27, no. 4, pp. 911–927, 2013.
[73] A. Biswas, P. Sahu, and M. Chandra, “Multiple cameras audio visual speech recognition using active appearance model visual features in car environment,” *International Journal of Speech Technology*, vol. 19, no. 1, pp. 159–171, 2016.
[74] F. Faubel, M. Georges, K. Kumatani, A. Bruhn, and D. Klakow, “Improving hands-free speech recognition in a car through audio-visual voice activity detection,” in *Proc. Joint Workshop on Hands-Free Speech Communication and.Microphone Arrays*, 2011.
[75] P. C. Loizou, *Speech Enhancement: Theory and Practice*, 2nd ed., Boca Raton, FL, USA: CRC, 2013.
[76] G. Hinton, N. Srivastava, and K. Swersky, “Lecture 6: Overview of mini-batch gradient descent,” Coursera Lecture slides https://class.coursera.org/neuralnets-2012-001/lecture.
[77] F. Chollet. (2015). *Keras*. Available: https://github.com/fchollet/keras
[78] J. W. Tukey, “Comparing individual means in the analysis of variance,” *Biometrics*, vol. 5, no. 2, pp. 99–114, 1949.
[79] C. Feichtenhofer, A. Pinz, and A. Zisserman, “Convolutional two-stream network fusion for video action recognition,” in *Proc. CVPR*, 2016.
[80] J. S. Chung, A. Senior, O. Vinyals, and A. Zisserman, “Lip reading sentences in the wild,” arXiv:1611.05358, 2016.
[81] S. Ntalampiras, T. Ganchev, I. Potamitis, and N. Fakotakis, “Objective comparison of speech enhancement algorithms under real world conditions,” in *Proc. PETRA*, 2008.
[82] S. Thermos, and G. Potamianos, “Audio-visual speech activity detection in a two-speaker scenario incorporating depth information from a profile or frontal view,” in *Proc. SLT, 2016.*
[83] F. Patrona, A. Iosifidis, A. Tefas, N. Nikolaidis, and I. Pitas, “Visual voice activity detection in the wild,” *IEEE Transactions on Multimedia*, vol. 18, no.6, pp. 967–977, 2016.
[84] J. Long, E. Shelhamer, and T. Darrell, “Fully convolutional networks for semantic segmentation,” in *Proc. CVPR*, 2015.
[85] S.-W. Fu, Y. Tsao, X. Lu, and H. Kawai, “Raw waveform-based speech enhancement by fully convolutional networks,” arXiv: 1703.02205, 2017.
[86] S.-W. Fu, Y. Tsao, X. Lu, and H. Kawai, “End-to-end waveform utterance enhancement for direct evaluation metrics optimization by fully convolutional neural networks,” arXiv:1709.03658, 2017.
[87] O. Ronneberger, P. Fischer, and T. Brox, “U-Net: convolutional networks for biomedical image segmentation,” in *Proc. MICCAI*, 2015.